%% file: ms.tex
\documentclass[useAMS,usenatbib]{mn2e}
\pdfoutput=0

\usepackage{aasmacros}
\usepackage{color}
\usepackage{epsfig}

\input{newcommands.tex}

\def\dn{{\rm D}_n4000}

\def\rhoc{\rho_{\rm crit}}
\def\rhob{\rho/\bar{\rho}}

\def\m12{M_{12}}

\def\m12{M_{12}}

\def\rs0{\hat{R}_{\rm sh}^0}
\def\mg2{Mg\,II}

\def\rhobar{\bar{\rho}}
\def\lstar{L_\ast}

\def\hcubed{(h^{-1}\,{\rm Mpc})^{-3}}

\def\wp{w_p(r_p)}

\def\fq{f_{\rm Q}}
\def\fold{f_{\rm old}}
\def\fqsat{f_{\rm Q}^{\rm sat}}
\def\fqcen{f_{\rm Q}^{\rm cen}}

\def\msub{M_{\rm sub}}
\def\mhost{M_{\rm host}}

\def\mgal{M_\ast}
\def\mhalo{M_{\rm halo}}

\def\fqsatobs{f_{\rm Q,obs}^{(\rm sat)}}
\def\fqcenobs{f_{\rm Q,obs}^{(\rm cen)}}
\def\rhonorm{\rho/\bar{\rho}}
\def\hhmsol{M_\odot/h^2}
\def\lesssim{\la}
\def\gtrsim{\ga}
\def\ntot{n_{\rm tot}}

\title[Halo and Galaxy Formation Histories]{Are Halo and Galaxy Formation Histories Correlated?}

\author[Tinker, Wetzel \& Conroy]{Jeremy L. Tinker$^1$, Andrew R. Wetzel$^2$, and Charlie Conroy$^3$\\
  $^1$Center for Cosmology and Particle Physics, Department of Physics, New York University, New York, NY\\
  $^2$ Department of Astronomy, Yale University, New Haven, CT\\
$^3$ Harvard-Smithsonian Center for Astrophysics, Combridge, MA}
\begin{document}


\pagerange{\pageref{firstpage}--\pageref{lastpage}} \pubyear{2011}

\maketitle

\label{firstpage}

\begin{abstract}

  The properties of dark matter halos, including mass growth,
  correlate with larger scale environment at fixed mass, an effect
  known as assembly bias.  However, whether this environmental
  dependence manifests itself in galaxy properties remains unclear.
  We apply a group-finding algorithm to Data Release 7 of the Sloan
  Digital Sky Survey to estimate the dark matter halo mass of each
  galaxy and to decompose galaxies into those that exist at the
  centers of distinct halos and those that orbit as satellites within
  larger halos. Using the 4000-\AA\ break as a measure of star
  formation history, we examine the correlation between the quenched
  fraction of galaxies, $\fq$, and large-scale environment, $\rho$. At
  all galaxy magnitudes, there is a positive, monotonic relationship
  between $\fq$ and $\rho$. We use the group catalog to decompose this
  correlation into the contribution from central and satellite
  galaxies as a function of halo mass. Because satellites are more
  likely to be quenched than central galaxies, the observed
  $\fq$-$\rho$ correlation is primarily due to variations of the halo
  mass function with environment, which causes a larger fraction of
  satellite galaxies at high $\rho$. For high-mass central galaxies,
  there is a weak but positive correlation between $\fq$ and $\rho$ at
  both fixed galaxy luminosity and fixed halo mass. For low-mass
  central galaxies ($\mgal\la 10^{10.0} \hhmsol$), there is no
  correlation between $\fq$ and $\rho$. The latter results are
  inconsistent with the strong assembly bias of dark matter halos seen
  in this mass regime if recent galaxy growth at all correlates with
  recent halo growth, as we demonstrate through a high-resolution
  $N$-body simulation. We also find that the mean stellar age of
  quenched central galaxies is independent of $\rho$ at fixed $\mgal$,
  while the formation times of low mass halos vary significantly. We
  conclude that the processes that halt the star formation of low mass
  central galaxies are not correlated to the formation histories of
  their host halos, and old galaxies do not reside preferentially in
  old halos.
\end{abstract}


\section{Introduction}

Many galaxy properties are correlated with their large- and
small-scale environment. The primary examples are the color-density
relation and the morphology-density relation (e.g.,
\citealt{oemler:74, davis_geller:76, dressler:80}). These results
demonstrated that, in the local Universe, galaxies in lower-density
regions are typically less massive, and that even at fixed luminosity
or mass, they are bluer and more likely to be spirals.  With the
advent of large-scale galaxy redshift surveys such as the Sloan
Digital Sky Survey (SDSS; \citealt{york_etal:00}), these seminal works
have been refined with increased statistical precision
(\citealt{hogg_etal:04, kauffmann_etal:04, blanton_etal:05a,
  baldry_etal:06, park_etal:07, bamford_etal:09}) and extended via
direct measurements of the correlations between star formation rates
and environment (\citealt{balogh_etal:04, kauffmann_etal:04}) and via
measurements out to $z\approx 1$ (\citealt{cucciati_etal:06,
  cooper_etal:07}).

Studies that investigated the relative importance of environment on
different scales indicated that small-scale environment (Mpc scale)
correlates stronger with galaxy properties than large-scale
environment (\citealt{kauffmann_etal:04, blanton_etal:06a,
  blanton_berlind:07, wilman_etal:10}). These results suggest that the
mass of the dark matter halo in which a galaxy resides plays the
dominant role in the galaxy's formation and evolution.  For example,
\cite{blanton_berlind:07} found that the blue fraction of galaxies in
rich groups and clusters is independent of environment at fixed halo
mass.  In this paper we extend these results by focusing on galaxies
in low-mass halos. 

The definition of environment is thus key to any discussion of this
topic. In this paper as in many other works, `environment' is defined
as the local \textit{galaxy} density at some smoothing scale.
However, the physical interpretation of this definition of environment
is not entirely straightforward because galaxy density is a biased
measure of dark matter density, in addition to being subject to shot
noise and the effects of redshift-space distortions (see, e.g.,
\citealt{cooper_etal:05}).  The choice of scale is also important. In
this paper we make the distinction between (large-scale) environment
and halo mass. We define environment as the local galaxy density on
scales larger than the host dark matter halo of a given galaxy, using
a fiducial smoothing scale of 10 \hmpc. Halo mass is similar to
`small-scale environment', but the distinction is important. The
dispersion of host halo masses for galaxies at a fixed 1 Mpc local
density can be 1 dex or more and using distance to the $n$-th nearest
neighbor yields an even larger spread in halo mass
\citep{haas_etal:11}.  This scatter demonstrates the need for
quantifying halo mass explicitly.

The idea that galaxy properties depend primarily on host halo
properties is physically motivated, since the halo virial radius
corresponds to a physical transition between infall and virialized
motions, capable of supporting strong shock fronts and hot,
thermalized gas \citep[e.g.,][]{dekel_birnboim:06}.  In this sense, a
galaxy group necessarily corresponds to a single, virialized dark
matter halo.  This is distinct from many other definitions of `group
environment'.  For example, the `Local Group' is not a group at all in
this context, but rather a close pair of separate halos. The
difference is an important one: for a halo to enclose both the Milky
Way and Andromeda (with one galaxy at the halo center), it would have
to be larger than $10^{13}$ \msol to equal their separation of $\sim
800$ kpc, a mass at which diffuse halo gas is heated to high
temperatures, causing shock heating and stripping of the disks in our
galaxy or M31 (or both). As a close pair of $\sim 10^{12}$ \msol\
halos, their tidal interactions and gas physics is markedly different:
there is no ram pressure and tidal stripping is much weaker.  It is
thus more informative to quantify halo mass rather than galaxy density
on a small scale.

However, environment does play a role in the formation of dark matter
halos.  In the numerical Universe, dark matter halos exhibit `assembly
bias': at fixed mass, halo properties depend on their formation
histories (\citealt{gao_etal:05, harker_etal:06, gao_white:06,
  wechsler_etal:06, wetzel_etal:07, li_etal:08, dalal_etal:08}). At
low halo mass, older and more concentrated halos form in high density
environments. At high mass the effect reverses, with younger, less
concentrated halos forming in high-density regions, though the effect
is also weaker.  Semi-analytic models of galaxy formation predict that
this assembly bias propagates into the evolution of galaxies at fixed
halo mass, especially in the formation histories and star formation
properties of low-luminosity galaxies in low-mass halos
(\citealt{zhu_etal:06, croton_etal:07}). At first glance, these
predictions appear in concert with the above observational evidence
that the fraction of galaxies (at fixed luminosity) that are red and
quenched is highest at high densities.

Another approach to understanding the connection between galaxies and
dark matter is the Halo Occupation Distribution (HOD; see, e.g.,
\citealt{seljak:00, peacock_smith:00, roman_etal:01,
  berlind_weinberg:02, cooray_sheth:02}). The HOD has emerged as our
most powerful tool for understanding the clustering of galaxies and
quantifying the bias between galaxies and dark matter, where bias here
is defined as the ratio between the galaxy and matter densities on an
arbitrary smoothing scale. The central quantity in the HOD is
$P(N|M)$, the probability that a halo of mass $M$ contains $N$
galaxies within a defined sample. The galaxy sample may be arbitrarily
defined, thus $P(N|M)$ will be different for every sample
definition. The bias of galaxies depends strongly on galaxy
properties, with strong clustering for galaxies that are brighter,
redder, and more elliptical (e.g., \citealt{norberg_etal:01,
  norberg_etal:02, zehavi_etal:02, zehavi_etal:05, zehavi_etal:10,
  li_etal:06}). In the HOD context, these trends are all explained by
the different halos that different types of galaxies occupy. Brighter
central galaxies live in more massive halos that are more highly
clustered. At fixed galaxy mass, redder or more elliptical galaxies
are more likely to be satellites in groups or clusters, thus enhancing
their clustering.  These observations fit into the HOD formalism
without any explicit correlation between galaxy formation and
environment, other than the fact that massive halos form at high
densities (e.g.~\citealt{bond_etal:91}).

The $P(N|M)$ ansatz has been extended even further in the `subhalo
abundance matching' approach, which (in its simplest form) assumes a
monotonic relationship between halo mass, $M$, and galaxy luminosity,
$L$ (\citealt{kravtsov_etal:04, wang_etal:06, vale_ostriker:06,
  conroy_etal:06, moster_etal:09, conroy_wechsler:09, wetzel_white:10,
  behroozi_etal:10}).\footnote{One can use halo maximum circular
  velocity rather than halo mass, or use galaxy stellar mass rather
  than galaxy luminosity.}  Here, if the number density of halos above
$M$ is the same as that of galaxies above $L$, then $M$ halos host
galaxies of $L$ brightness. They key to this approach is to include
not just isolated halos, but also `subhalos', defined as bound,
virialized halos located within the virial radii of larger host
halos. The primary assumption of this approach is that (sub)halo mass
entirely determines the luminosity of the galaxy within it, regardless
of any other properties of the (sub)halo or the galaxy.

The quantity $P(N|M)$ makes the explicit, and strong, assumption that
$N$ depends only on $M$.  That is, the properties of galaxies,
including luminosity, color, age, and morphology, depend only on the
mass of their host dark matter halo and have no additional dependence
on a halo's formation history or environment.  This formulation was
not proposed with any theoretical prejudice about galaxy formation,
but rather it was the simplest implementation and was successful in
fitting the clustering properties of nearly all available
data. However, the discovery of halo assembly bias, combined with the
complex relation between galaxy properties and environment, especially
those governing galaxy growth and star formation, calls into question
this ansatz.

In \cite{tinker_etal:08_voids}, we sought to test this $P(N|M)$ ansatz
through different galaxy clustering measures that probe different
environments. We used the two-point correlation function, $\xi_g(r)$,
to constrain halo occupation and make predictions for the void
probability function. The predictions were in excellent agreement with
the measurements, demonstrating the validity of the HOD from high to
low densities. \cite{abbas_sheth:06} and \cite{skibba_etal:06} used
other galaxy clustering measures to demonstrate that the assumption of
$P(N|M)$ is consistent with observed data for luminosity-based galaxy
samples. However, \cite{yang_etal:06} and
\cite{wang_etal:08_assembly_bias} found that the clustering of
low-mass galaxy groups at fixed mass depends on the color of the
galaxies contained within them such that groups with redder galaxies
are more highly clustered.

In this paper, we examine the relation between galaxy star formation
histories, their host dark matter halos, and their large-scale
environment.  We explore in particular the extent to which the
formation histories of galaxies relate to those of their host halos
and whether the assembly bias seen in simulation extends to the galaxy
population.  We focus on low-mass galaxies, the regime in which halo
assembly bias effects are the strongest, and we examine star formation
history based on narrow 4000-\AA\ break, $\dn$, which, unlike color,
is insensitive to dust reddening.  In order to observationally
determine halo masses for SDSS galaxies and separate this from
large-scale environment, we construct galaxy group catalogs using the
\cite{yang_etal:05} group-finding algorithm.  Our use of a group
finder is critical because galaxies of the same properties may occupy
halos of a wide range of mass. At fixed stellar mass, galaxies may
live at the center of low-mass halos or they may exist as satellites
in group or cluster-sized halos. Anywhere between 10-40\% of galaxies
are satellites, depending on galaxy mass and color (see, e.g.,
\citealt{yang_etal:08, zehavi_etal:10}). The distinction between
central and satellite galaxies will be the cornerstone of
understanding the environmental correlations of galaxy properties.

The evolution of satellite galaxies is more complex, and their
properties also depend on position within a halo (e.g.,
\citealt{vdb_etal:08b,hansen_etal:09,weinmann_etal:10}). Here we focus
on the properties of central galaxies in halos below the group
scale. We will explore the observed dependence of satellite star
formation rate on galaxy mass, halo mass, and halo-centric radius in
detail in a companion paper (Wetzel, Tinker, \& Conroy~2011a),
hereafter referred to as Paper II.  Finally, in a third paper (Wetzel,
Tinker, \& Conroy~2011b), hereafter referred to as Paper III, we will
use these measurements combined with a high-resolution simulation to
test various physical mechanisms for the quenching of satellite
galaxies.

We examine galaxies selected on either $r$-band magnitude or stellar
mass, $\mgal$, and we examine star formation history based on $\dn$.
Magnitude and $\dn$ have the advantage of being well-defined
observationally, though in Paper II we will focus entirely on stellar
mass and instantaneous specific star formation rate (SSFR), quantities
which are derived and thus more model-dependent, but also more readily
physically interpretable.  We have examined all halo and environmental
trends in both papers selecting on both $M_r$ and $\mgal$ as well as
examining both $\dn$ and SSFR.  While these lead to slight
quantitative differences in our results, they do not change any of our
results qualitatively.

Throughout, we define a galaxy group as any set of galaxies that
occupy a common dark matter halo (including cluster-mass halos), and
we define a halo as having a mean interior density 200 times the
background matter density. A {\it host halo} is a halo that is
distinct: its center does not reside within the radius of a larger
halo. A {\it subhalo} is one whose center is located within the radius
of a larger halo. For all calculations we assume a flat, \lcdm\
cosmology of $(\om,\s8,\omb,n_s,h_0)=(0.27, 0.82,
0.045,0.95,0.7)$. For galaxy magnitudes we do not assume a value of
$h$, but for brevity we write all magnitudes as $M_r$ rather than
$M_r-5\log h_0$. Stellar masses are in units of $\hhmsol$.

\begin{figure*}
\centering
\begin{tabular}{cc}
\psfig{file=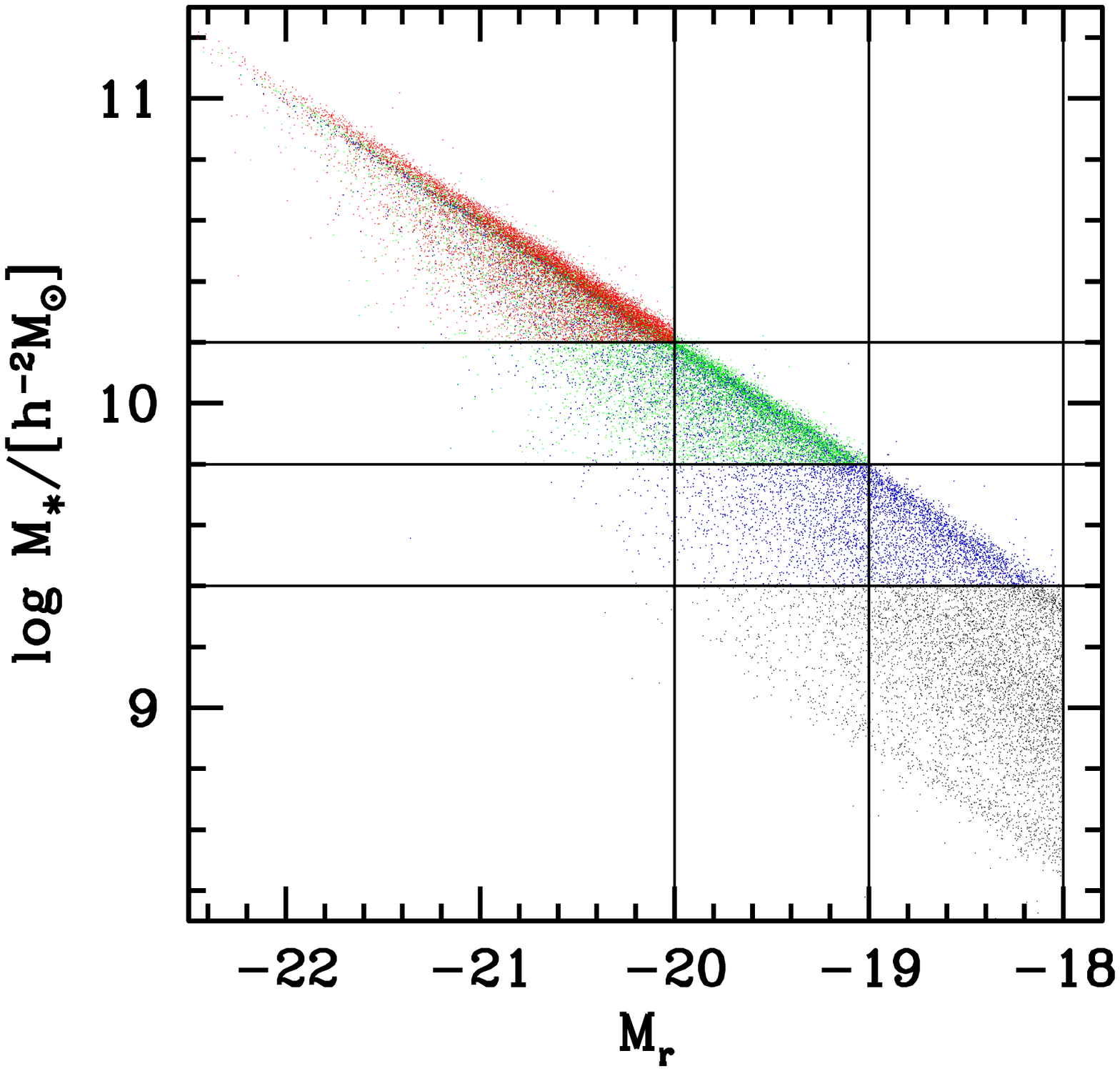,width=0.5\linewidth,clip=} &
\psfig{file=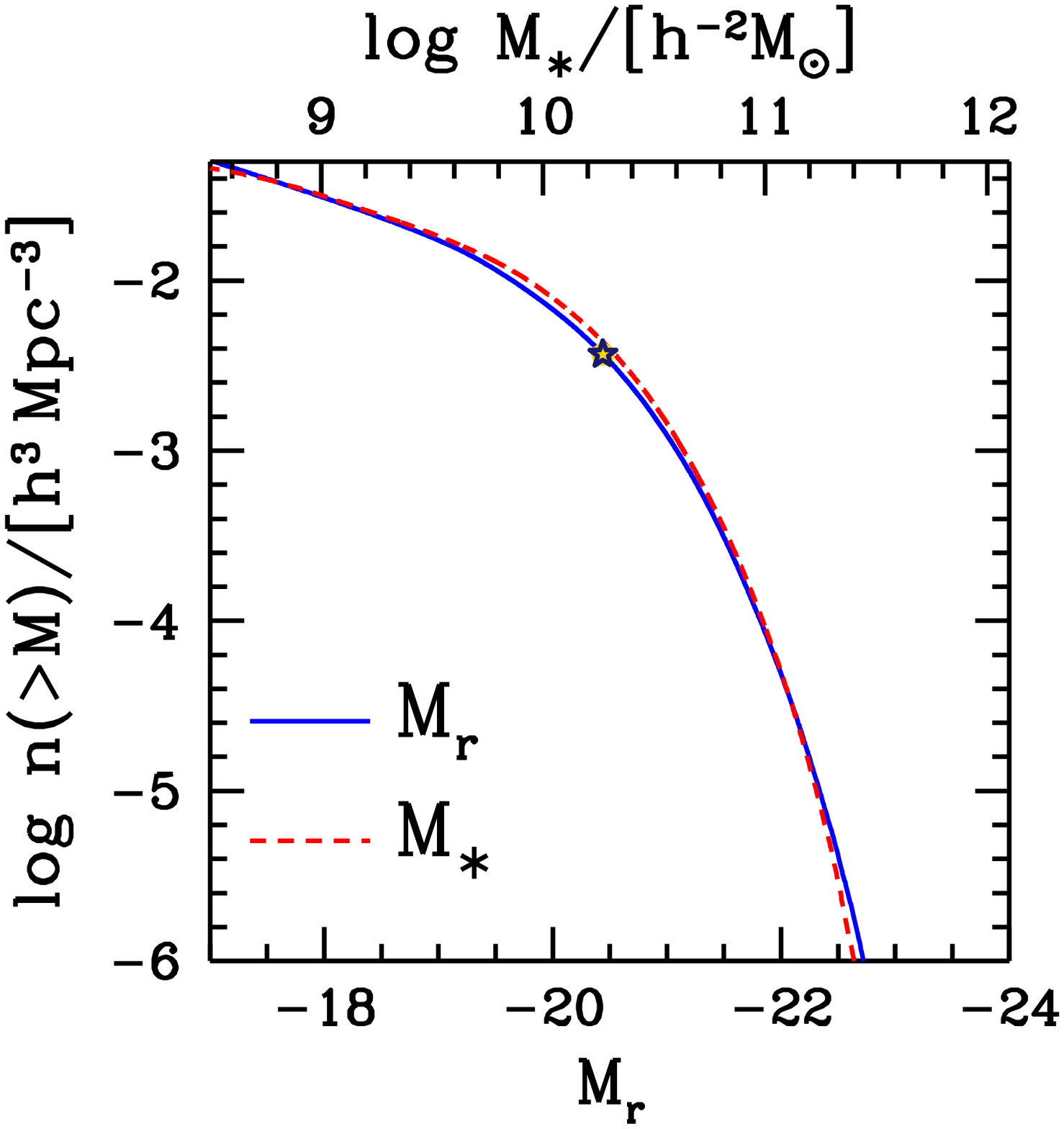,width=0.5\linewidth,clip=} 
\end{tabular}
\caption{ \label{samples} {\it Left panel}: Relation between $r$-band
  magnitude, $M_r$, and stellar mass, $\mgal$, with the latter taken 
  from the {\tt kcorrect} code of
  \citet{blanton_roweis:07}. Vertical lines show the magnitude
  limits for our volume-limited samples as listed in Table 1, while
  horizontal lines indicate the stellar mass limits within each
  volume-limited sample. {\it Right panel}: Cumulative number density
  of galaxies as a function of both $M_r$ and $\mgal$, offering 
	a rough conversion between the two.
	The star indicates $M_r^\ast$ from the \citet{blanton_etal:03} 
  luminosity function.}
\end{figure*}


\begin{table}
\centering
\begin{minipage}{140mm}
\caption{Volume-Limited Samples}
\begin{tabular}{@{}ccccc@{}}
\hline
$M_r$ & $z_{\rm max}$ & $N_{\rm gal}(<M_r)$ &  $\log M_\ast/[h^{-2}\,M_\odot]$ & $N_{\rm gal}(>M_\ast)$\\
\hline
-18.0 & 0.040 & 31729 & 9.4 & 21423 \\
-19.0 & 0.064 & 74987 & 9.8 & 54119 \\
-20.0 & 0.103 & 131658 & 10.2 & 100852 \\
\end{tabular}
\end{minipage}
\end{table}

\begin{figure}
\centerline{\psfig{figure=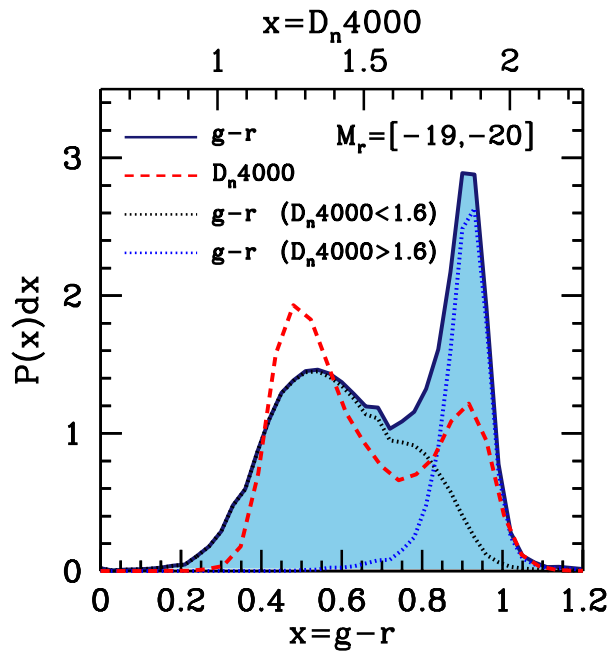,width=9.0cm}}
\caption{ \label{color_dn4k} The distribution of $g-r$ color and $\dn$
  for $M_r=[-19,-20]$ galaxies. The shaded region shows the
  distribution of color, with scale on the bottom $x$-axis. The red
  dashed curve shows the distribution of $\dn$ for the same galaxies,
  with scale on the top $x$-axis. Even though 53\% of galaxies are
  red $(g-r>0.7)$, only 35\% of galaxies are truly quenched
  $(\dn>1.6)$. The blue and black dotted curves show the color
  distribution for quenched and active galaxies, respectively,
  according to their $\dn$ values. Quenched galaxies nearly always
  have red colors, while 23\% of active galaxies have red colors, 
  primarily because of dust extinction (e.g., \citealt{maller_etal:09}).}
\end{figure}

\begin{figure}
\centerline{\psfig{figure=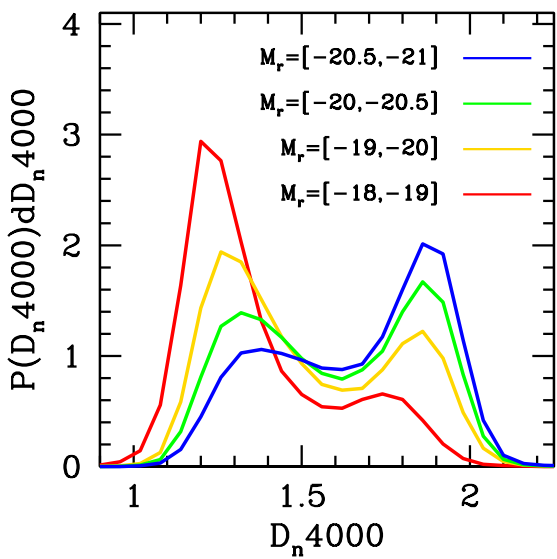,width=8.0cm}}
\caption{ \label{galprop_histo} Distribution of $\dn$
  values for galaxies in magnitude bins. Faint galaxies are
  predominantly active while bright galaxies are predominantly
  quenched but still have a significant active population. All
  sample distributions are bimodal with a minimum at $\dn \approx 1.6$.
	We refer to galaxies with $\dn > 1.6$ as `quenched'.}
\end{figure}

\begin{figure}
\centerline{\psfig{figure=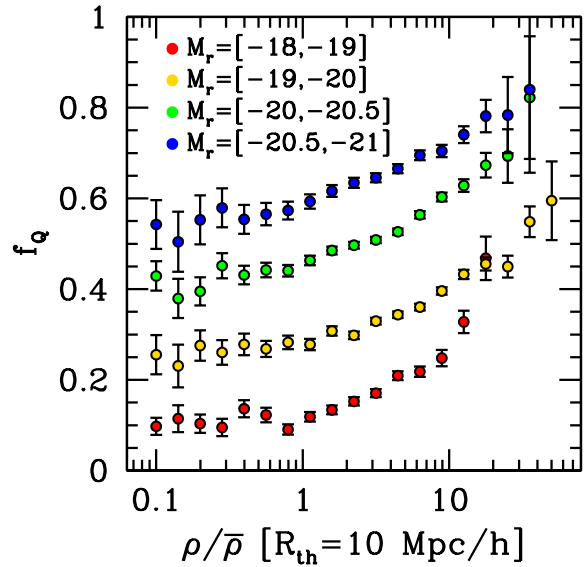,width=8.0cm}}
\caption{ \label{galprop_density} Quenched fraction, $\fq$, defined by
  $\dn > 1.6$, as a function of large-scale (10 \hmpc) density of galaxies. At
  $\rhob>1$, $\fq$ rises monotonically with density, while in
  underdense regions the quenched fraction is nearly constant.}
\end{figure}

\section{Data and Measurements}

\subsection{NYU Value-Added Galaxy Catalog}

To construct our galaxy samples, we use the NYU Value-Added Galaxy 
Catalog (VAGC; \citealt{blanton_etal:05_vagc}) based on the spectroscopic 
sample in Data Release 7 (DR7) of the Sloan Digital Sky Survey 
(SDSS; \citealt{dr7}). Specifically we use \texttt{dr72bight34}. 
We construct three volume-limited samples, as listed in Table 1, which 
contain all galaxies brighter than $M_r=-18$, $M_r=-19$ and $M_r=-20$, 
respectively. Within each volume-limited sample, we construct another 
sample that is complete in stellar mass. The stellar masses are
also taken from the VAGC and are derived from the {\tt kcorrect} code
of \cite{blanton_roweis:07}, which assumes a \citet{chabrier:03} initial 
mass function. Fig.~\ref{samples} (left) shows the comparison
between stellar mass and $r$-band magnitude. The vertical and
horizontal lines indicate the magnitude and stellar mass limits,
respectively, for our samples. By constructing our stellar mass
samples this way, they contain significantly fewer galaxies than the
luminosity-defined samples. This limits the statistics and effective halo 
mass range of the stellar mass samples, but this
method has the benefit of requiring no weighting scheme for red
galaxies that are fainter than blue galaxies at fixed stellar mass. We
will construct groups and present results based upon both sample 
definitions in order to demonstrate robustness to galaxy property selection. 
To allow convenient, though approximate, conversion between the two 
properties, Fig.~\ref{samples} (right) also shows the cumulative number 
densities of galaxies based upon $M_r$ and $\mgal$.

For galaxy pairs that are too close to obtain spectra because of
the 55 arcsecond width of SDSS fibers (`fiber collisions'), we use the
internal correction to the fiber corrections within the VAGC, namely
that the collided object is given the redshift of the nearest galaxy
in terms of angular separation, provided that this redshift is in
agreement with the photometric redshift obtained by with the SDSS
photometry (\citealt{blanton_etal:05_vagc}).

\subsection{Bimodality of Galaxy Properties} 

Galaxies can be roughly divided into two distinct color categories:
the `blue cloud' and the `red sequence'. The former is comprised
of galaxies that are actively forming stars at the current epoch,
giving them blue colors. These galaxies are primarily disk-dominated
with significant amounts of cool gas. The red sequence is comprised
(mostly) of galaxies with old stellar populations, devoid of cold
gas, and usually exhibiting elliptical morphology. The bimodality of the
galaxy color distribution is well measured at $z=0$
(\citealt{strateva_etal:01, blanton_etal:03cmd, kauffmann_etal:03b,
  madgwick_etal:03}) and has been shown to exist at $z=1$
(\citealt{bell_etal:04, cooper_etal:06, willmer_etal:06}) and even up
to $z=2$ (\citealt{williams_etal:09}).

Using galaxy color as a proxy for star formation activity can be
problematic, as dust reddening can cause a gas-rich disk galaxy to be
classified as a red sequence object
\citep{maller_etal:09,masters_etal:10_dust}.  To avoid this problem,
we use $\dn$, which is derived from SDSS spectra. $\dn$ is a diagnostic
of the light-weighted age of the stellar population and thus is sensitive to the
integrated star formation history of the galaxy.  We obtain this
quantity from the JHU-MPA spectral reductions\footnote{\tt
  http://www.mpa-garching.mpg.de/SDSS/DR7/}
(\citealt{brinchmann_etal:04}). Fig.~\ref{color_dn4k} compares the
distribution of $g-r$ color to that of $\dn$ for $-19>M_r>-20$
galaxies, our fiducial magnitude bin throughout this paper.  The red
sequence is defined a strong peak in the color distribution at
$g-r\approx 0.9$ and at $\dn\approx 1.9$. Both distributions are
clearly bimodal, but based on color alone one would conclude that the
majority of galaxies in this magnitude range are quenched. The picture
painted by $\dn$ is strikingly different: the majority of galaxies of this
magnitude are instead active, but many have been dust-reddened.  We
have confirmed that this population of active but red galaxies is
caused primarily by dusty spirals by applying Galaxy Zoo morphological
classifications \citep{lintott_etal:11} to the sample: $\sim70\%$ of
are identified as spirals, and $\sim50\%$ are identified as edge-on
spirals \citep[see also][]{masters_etal:10_dust}.

One concern in using galaxy properties defined by spectra is fiber
collisions, which cause $\sim 8\%$ of all galaxies in the survey to
not have spectra. To correct for this, for each fiber-collided galaxy
we draw a random value of $\dn$ based on a sample of spectroscopic
galaxies that are within $\pm 0.25$ in $M_r$ and $\pm 0.05$ in $g-r$
of the collided object.  This method allows us to use the colors to
infer statistically the spectroscopic properties of the collided
sample. This method reflects the fact that not all collided red
galaxies are truly quenched, and it preserves the fact that the color
distribution of collided objects (and the distribution of $\dn$) is
distinct from the overall color distribution at the same
magnitude. Using this method, our results are generally consistent
with those obtained by simply removing the collided galaxies from the
samples after constructing the group catalog.

Fig.~\ref{galprop_histo} shows $\dn$ distribution for galaxies of
different absolute magnitudes. As seen in previous studies, the
bimodal distribution in sub-$\lstar$ galaxies is apparent but
is dominated by an younger, active population. At $L\gtrsim
\lstar$, the bimodality is still present but has been tilted to the
older, quenched population. The brightest galaxies are
dominated by an red, old population that is not forming stars at any
appreciable rate. Because the minimum of the bimodality 
occurs at $\dn = 1.6$ across all magnitudes we explore, we 
define galaxies with $\dn<1.6$ as `active' and $\dn>1.6$ as 
`quenched'.

Another possible concern with galaxy properties defined by SDSS
spectra is the finite 3 arcsec aperture of the fibers, corresponding
to $\sim2\,$\hkpc\ at the redshifts we consider. Thus, for many
galaxies the SDSS fiber covers only the inner light profile, where the
stellar population may by older (or younger) than the rest of the
galaxy. As can be seen in Fig.~\ref{galprop_histo}, the fraction of
blue galaxies that are labeled as quenched via $\dn$ is negligible
compared to the fraction of red galaxies that have active stellar
populations. Thus, while aperture effects may bias $\dn$ values
somewhat, the effect of dust obscuration on color is far more
important.

\subsection{Measuring Large-scale Environment}

For each galaxy, we determine the large-scale environment by counting
the number of neighboring galaxies within a sphere of radius 10 \hmpc\ 
centered on each galaxy.  This quantity is a biased indicator of the 
dark matter density field, but at 10 \hmpc\ this bias is a simple linear 
factor and any stochasticity is minimal. We count the number of galaxies 
above the corresponding magnitude threshold for the each sample, and so the 
tracer of the density field has a different bias for each sample. We do not
correct for this between the samples, but note that the relative bias
between $M_r<-18$, $-19$, and $-20$ samples is at the $\sim 5\%$ level
(\citealt{swanson_etal:08_bias}). This galaxy density measurement is affected 
by galaxy peculiar velocities, but this effect is minimal at 10 \hmpc, 
as we demonstrate in Appendix A.

We also choose 10 \hmpc\ as it represents a clear distinction from a
galaxy's small-scale environment as encapsulated by its host halo,
especially since 10 \hmpc\ represents the maximum extent from which
the most massive galaxy clusters have accreted their mass. In tests we
find that our results show little dependence on the exact smoothing
scale chosen. In Paper II we will further discuss the properties of
galaxies on the outskirts of groups and clusters.


To correct for survey geometry and incompleteness, we use of random catalogs. 
For each volume-limited sample, we produce a catalog of $10^7$ random
points distributed with the angular selection function of SDSS DR7
using the angular mask provided with the VAGC in combination with the
software package \texttt{mangle} (\citealt{swanson_etal:08}). Each
random point is also assigned a random redshift such that the comoving
space density of randoms is constant with redshift. For each galaxy,
we correct for incompleteness by multiplying the observed number of
galaxies by the ratio of the number of random points divided by the
expected number of randoms if the completeness were unity. The large
number of random points ensures that shot noise within each 10 \hmpc\
sphere is at the sub-percent level.

Fig.~\ref{galprop_density} shows the fraction of quenched galaxies as
a function of large-scale density. For all magnitude bins there is
clear dependence on density such that quenched galaxies preferentially
live in overdense environments. At $\rho/\bar{\rho}>1$, the slope of
the quenched fraction with $\log\rhob$ is roughly independent of
galaxy magnitude.  At $\rho/\bar{\rho}>10$, the fraction of quenched
galaxies in the faintest magnitude bin rises steeply. Other studies
have also noted that the faintest red galaxies predominantly are found
in the highest density regions (\citealt{hogg_etal:04,
  blanton_etal:05a}).

\begin{figure}
\centerline{\psfig{figure=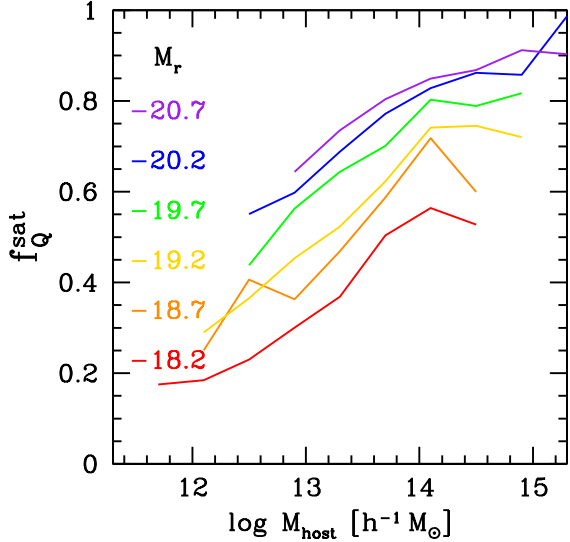,width=8.0cm}}
\caption{ \label{fquench_hostmass} Quenched fraction of
  satellite galaxies, $\fqsat$, as a function of $\mhost$ for bins in
  galaxy magnitude. At fixed halo mass, brighter satellites are more 
  likely to be quenched, and satellites of all magnitudes show a similar 
  increase in quenched fraction with halo mass.}
\end{figure}

\section{Group Finding Algorithm}

Motivated by the framework of the HOD model, we assign halos to
galaxies using the group-finding algorithm detailed in
\cite{yang_etal:05} and applied to the SDSS in a series of papers
(\citealt{yang_etal:07_catalog, yang_etal:08, yang_etal:09_groups3,
  weinmann_etal:06a, wang_etal:08_assembly_bias, vdb_etal:08}).  Given
the known statistics of dark matter halos, the halo masses of groups
of galaxies can be retrofit to the observed distribution of
galaxies. Because this algorithm has been described in detail in the
papers above, in this section we give a synopsis of the group finder
and the minor modifications we made to the algorithm, and we provide a
more thorough description in Appendix B.

Our starting point differs from the algorithm of
\cite{yang_etal:05}. To begin, we ascribe to each galaxy in the VAGC a
dark matter halo mass via the subhalo abundance matching method
described in \S 1. Typically, abundance matching is used to put
galaxies into a population of dark matter (sub)halos, but we invert
this process to put {\it (sub)halos} around {\it galaxies} as the
first step in the group-finding algorithm. The abundance matching
method relates galaxy luminosity (or stellar mass) to (sub)halo mass
by assuming that the two are monotonically related. Thus, a (sub)halo
of mass $M_0$ contains a galaxy of luminosity $L_0$ such that

\begin{equation}
\label{e.sham}
\int_{L_0}^\infty \Phi(L)\,dL = \int_{M_0}^\infty \ntot(M)\,dM
\end{equation}

\noindent where $\Phi(L)$ is the galaxy luminosity function (or
alternately the stellar mass function) and $\ntot(M)$ is the halo plus
subhalo mass function.

In applying equation (\ref{e.sham}) we use the \cite{tinker_etal:08_mf}
function for the halo mass function and the subhalo mass function in
\cite{tinker_wetzel:10} (their equation 12; but we have changed the 
normalization from 0.3 to 0.2 to account for the differences in the
subhalo mass function between redshift 1 and 0). The total number density 
of halos of mass $M$ is then

\begin{equation}
\label{e.mf_smf}
\ntot(M) = n(M_{\rm host}) + 
\int n(\msub |M_{\rm host}) n(M_{\rm host}) \,dM_{\rm host}
\end{equation}

\noindent where we have denoted halos that are distinct as `host'
halos and halos that are contained within host halos as `sub'.
Thus, each galaxy is given a halo mass regardless of whether it is a
satellite within a larger host halo or resides in the
field\footnote{Note that $\msub$ is the mass of the subhalo at the
  time of accretion. See details in \cite{tinker_wetzel:10} and a
  discussion of subhalo finding in Paper III.}.

\begin{figure*}
\centerline{\psfig{figure=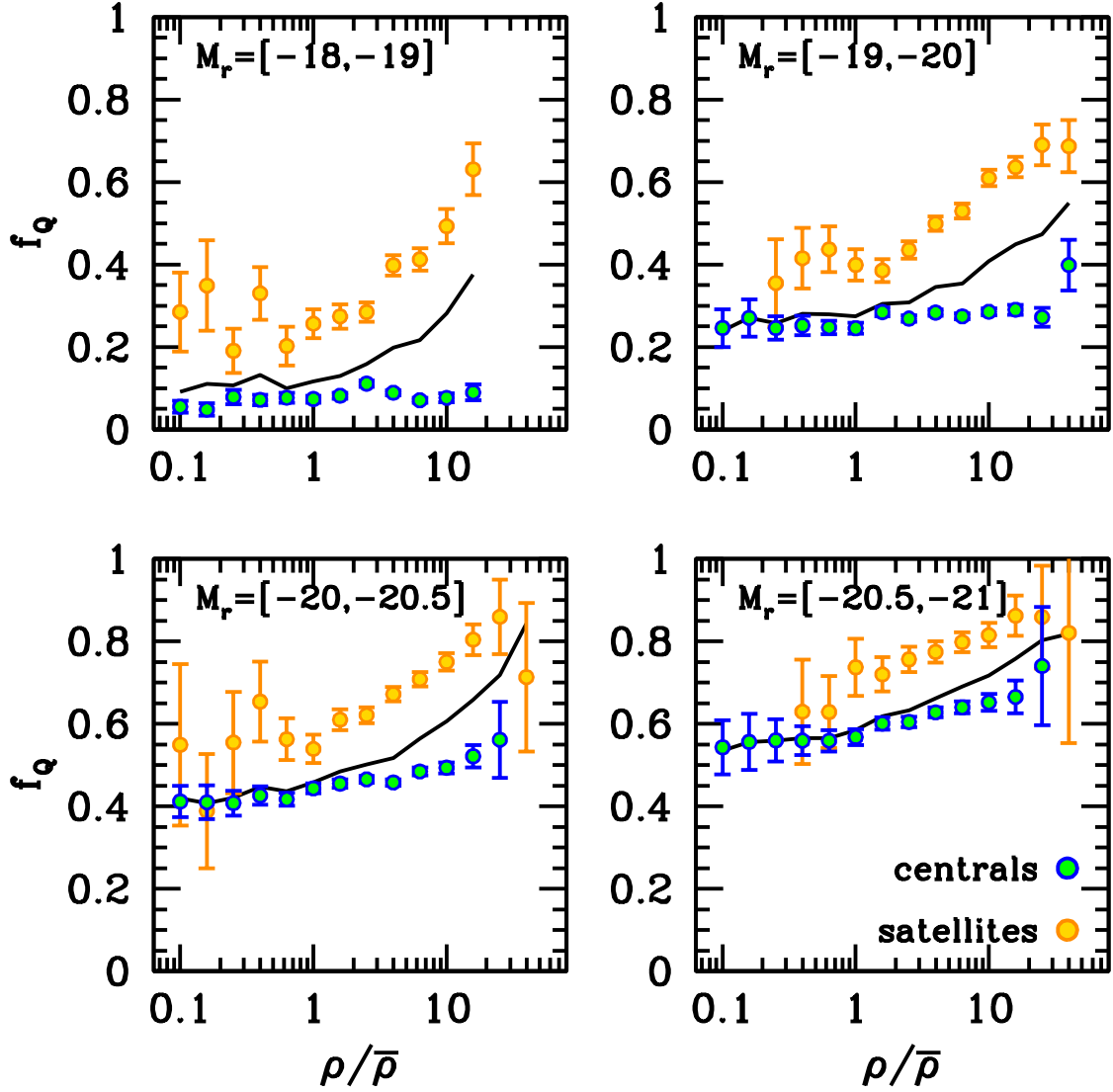,width=16.0cm}}
\caption{ \label{fq_den_4win} Quenched fraction, $\fq$, vs. 10 \hmpc\
  overdensity, $\rhob$, for various bins in galaxy magnitude. In each
  panel, the solid curve is the overall $\fq$-$\rhob$ relation, with
  the filled circles showing $\fqcen$ and $\fqsat$.  Bright central
  galaxies do show a slight increase in quenched fraction with
  density.}
\end{figure*}

\begin{figure*}
\centerline{\psfig{figure=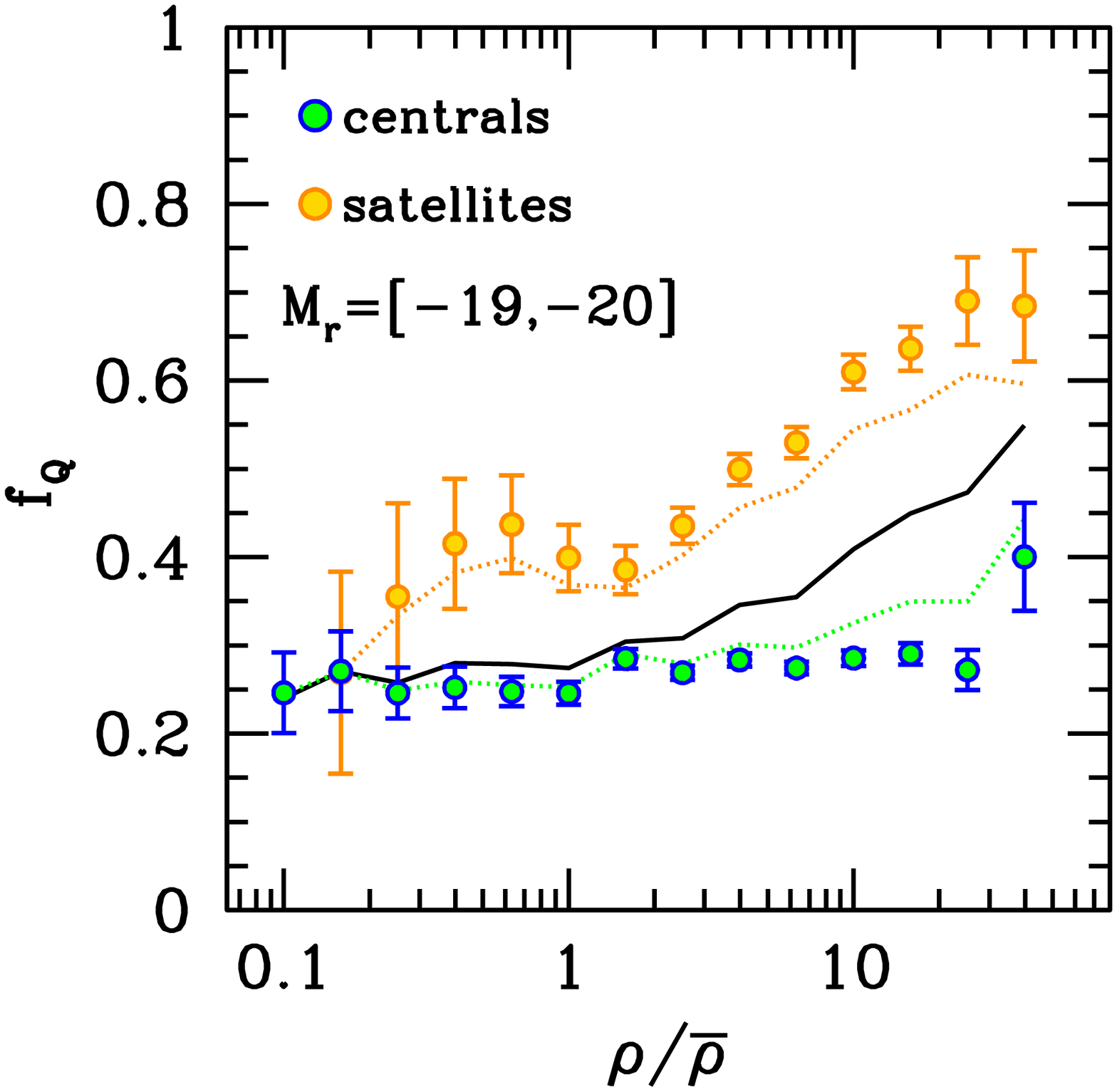,width=8.0cm} \psfig{figure=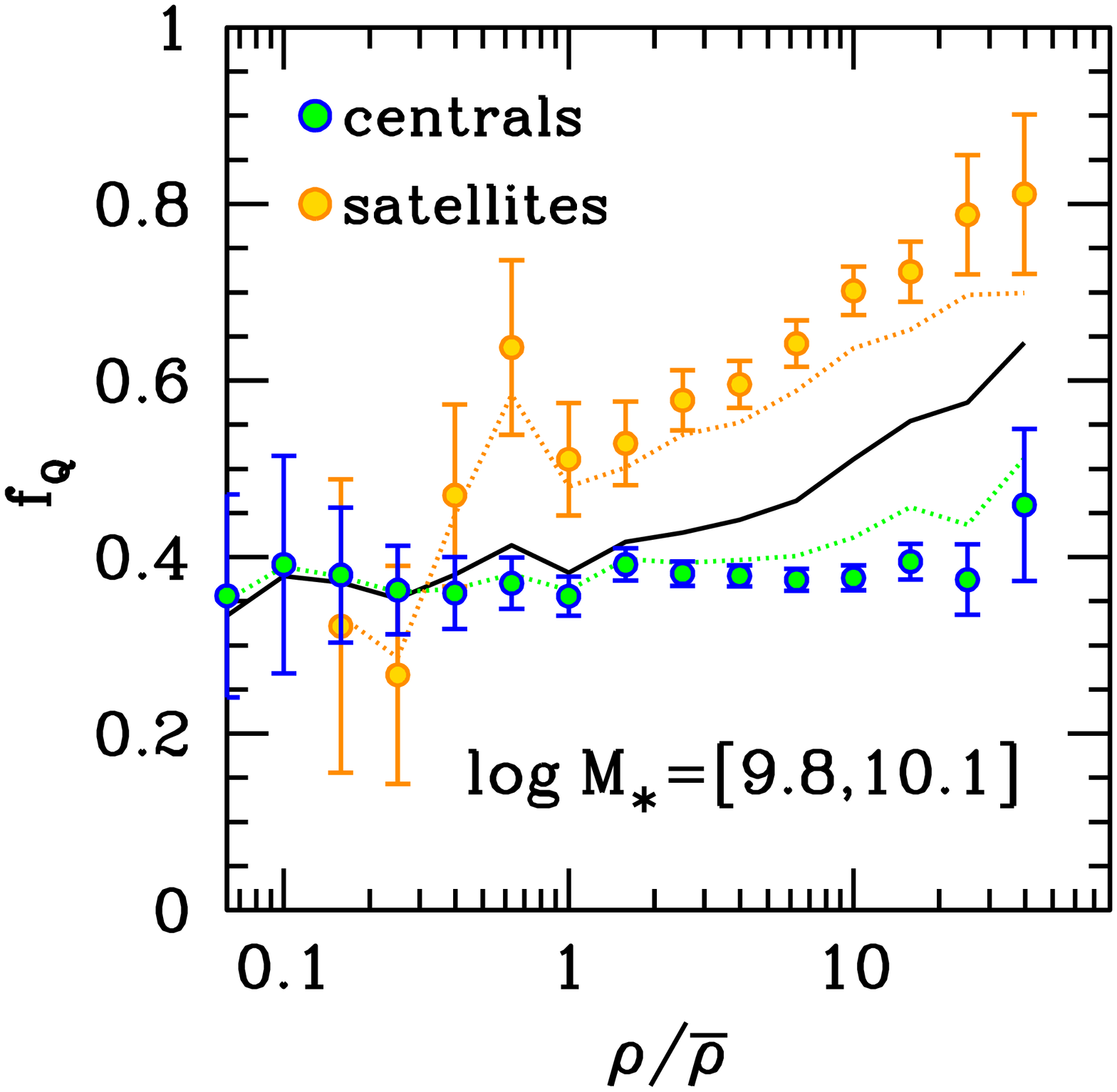,width=8.0cm}}
\caption{ \label{fq_den_decomp} {\it Left panel:} The relation between
  quenched fraction, $\fq$, and 10 \hmpc\ overdensity, $\rhob$, for
  $M_r=[-19,-20]$ galaxies. The solid curve shows the overall relation
  from Fig.~\ref{galprop_density}. The different filled circles show
  the quenched fraction divided into central and satellite
  galaxies. The dotted curves show the measurements without correcting
  for satellite-central misclassification in the group finder (this
  correction is described in detail in Appendix C). {\it Right panel:}
  Same as left panel, but using the galaxy group catalog defined with
  stellar mass as opposed to galaxy luminosity as the tracer of group
  mass. The number density of galaxies with $\mgal>10^{9.8} \hhmsol$
  is $\sim 30\%$ lower than that of galaxies with $M_r<-19$ (see
  Fig.~\ref{samples}), thus the central galaxies in the right hand panel
  occupy slightly higher mass halos. }
\end{figure*}

To obtain the left-hand side of equation (\ref{e.sham}), we sum all
galaxies, $L_i$, brighter than $L_0$, weighed by $V_{\rm max}$ to
obtain the number density, given by,

\begin{equation}
\label{e.lumfun}
n(>L_0) = \sum_{L_i>=L_0} V^{-1}_{\rm max}(L_i)
\end{equation}

\noindent where $V_{\rm max}$ is the maximum volume to which each
galaxy is observed, as calculated in the VAGC to account for survey
incompleteness. Once again, luminosity can be replaced by stellar
mass, and a comparison of the cumulative density of galaxies as a
function of magnitude and $\mgal$ is given in Fig.~\ref{samples}.
Although we later cull the VAGC to create the volume-limited samples
in Table 1, using the full flux-limited catalog suppresses sample
variance in the calculation of the luminosity function for the initial
halo mass assignment.

In the \cite{yang_etal:05} methodology, this initial matching of
galaxies to groups and halo mass is done with a redshift-space
friends-of-friends (FOF) linking algorithm (see also
\citealt{berlind_etal:06_catalog}). Using the inverse abundance
matching approach, we find results that are consistent with the FOF
algorithm.  See the tests in Appendix C.

Once a halo has been assigned to each galaxy, each galaxy has an
associated halo mass, virial radius, and velocity dispersion via the
virial theorem.  We then determine the probability that each galaxy is
a central galaxy in a host halo or a satellite galaxy in a subhalo.
If, projected on the sky, a galaxy lies within a more massive galaxy's
virial radius, we determine an angular probability that the
galaxy is a satellite by assuming that the number density profile of
satellite galaxies follows the dark matter given by an NFW
\citep{nfw:97} profile. We assume the concentration-mass relation
given by \cite{maccio_etal:08} for our cosmology, but note that the
results are insensitive to this choice. We also assign a line-of-sight
satellite probability to the galaxy given its redshift offset from the
more massive galaxy, where we assume that the host halo's satellites
are distributed in a Gaussian along the line of sight. If the product
of the angular and line-of-sight probability is above a calibrated
constant, then the galaxy is considered to be a satellite in the
larger host halo. Once we have applied this routine to all galaxies
and we have a list of candidate host halos, we recompute host halo
masses, again by the abundance matching method. However, from this
point on, we assume instead that host halo mass correlates with the
{\it total} luminosity of the group.  Thus, we use the halo mass
function in equation (\ref{e.sham}) for host halos only (no subhalos),
and the luminosity now is the {\it total} luminosity of the galaxies
within the halo. We iterate this procedure until host halo masses for
the groups have converged.  Our catalog contains groups varying in
mass from $2\times10^{11}$ to $10^{15}\,h^{-1}M_\odot$.

For each sample, we perform the group-finding procedure once using
total galaxy luminosity as the tracer of halo mass and once using
stellar mass as the tracer of halo mass. There is evidence that
stellar mass correlates stronger with halo mass than luminosity,
especially when dividing a galaxy sample by color
(\citealt{mandelbaum_etal:06_gals, more_etal:10}). However, there are
also significant uncertainties in determining stellar mass from
broadband optical magnitudes (e.g., \citealt{conroy_etal:09,
  conroy_gunn:10}), and so it is not clear whether stellar mass or
luminosity is preferable here.  Nonetheless, we will demonstrate that
all our conclusions are robust to choice of halo mass tracer.

Note that our algorithm defines the brightest (or most massive) group
galaxy as the central galaxy, that is, the galaxy residing at the
minimum of the host halo's potential well. For halos with
$\mhost\lesssim 10^{13}$ \hmsol, central galaxies are expected to be
the brightest galaxy in the halo the vast majority of the time
(\citealt{zehavi_etal:05, tinker_etal:05, zheng_etal:05,
  yang_etal:09_groups3}). For cluster-sized halos, $\mhost\gtrsim
10^{14}$ \hmsol, whether the brightest galaxy is always central galaxy
is not clear \citep{skibba_etal:11}. In statistical terms, halos of
this mass are rare and do not contribute significantly to the results
presented here. Because we focus on central galaxies with $L\la\lstar$
luminosities, this is not a concern for our results.

\cite{yang_etal:05} and subsequent papers have demonstrated the
efficacy of the halo-based group finder, but the approach is not
infallible. The algorithm is designed to obtain the correct mean
number of galaxies per halo, but because of projection effects and
redshift-space distortions, some galaxies will erroneously be labeled
as `central' when in fact they are within the virial radius of a
larger halo, while other central galaxies will falsely be labeled as
`satellites'.  This misclassification of galaxy type
means that the measured fraction of central and satellite galaxies
that are quenched will be biased because satellite galaxies are more
likely to be quenched than central galaxies at fixed $\mgal$. In
Appendix C we use a high-resolution $N$-body simulation to quantify the
false-classification rate and demonstrate how to statistically correct
the measurements of $\fq$ for this effect. On average, $\sim 10\%$ of
galaxies are misclassified, and all of the results in this paper
include a correction for this misclassification, except where stated
otherwise. We note that the group-finding algorithm is actually {\it
  more} efficient at identifying the central galaxies than satellite
galaxies within a given sample, yielding a completeness of $~\sim
95\%$ for most central galaxies (see App. C).

We have also checked our results against a SDSS DR7 version of the
\citet{yang_etal:07_catalog} group catalog (kindly supplied by Frank
van den Bosch), incorporating the same {\tt kcorrect} stellar masses $\dn$ 
measurements we use here but retaining differences
in assumed cosmology, halo virial definitions, and group-making
methodology.  All of the results are consistent within errors,
demonstrating the robustness of the iterative group-making method to
the details of the initialization of the procedure.

\section{Results}

\begin{figure}
\centerline{\psfig{figure=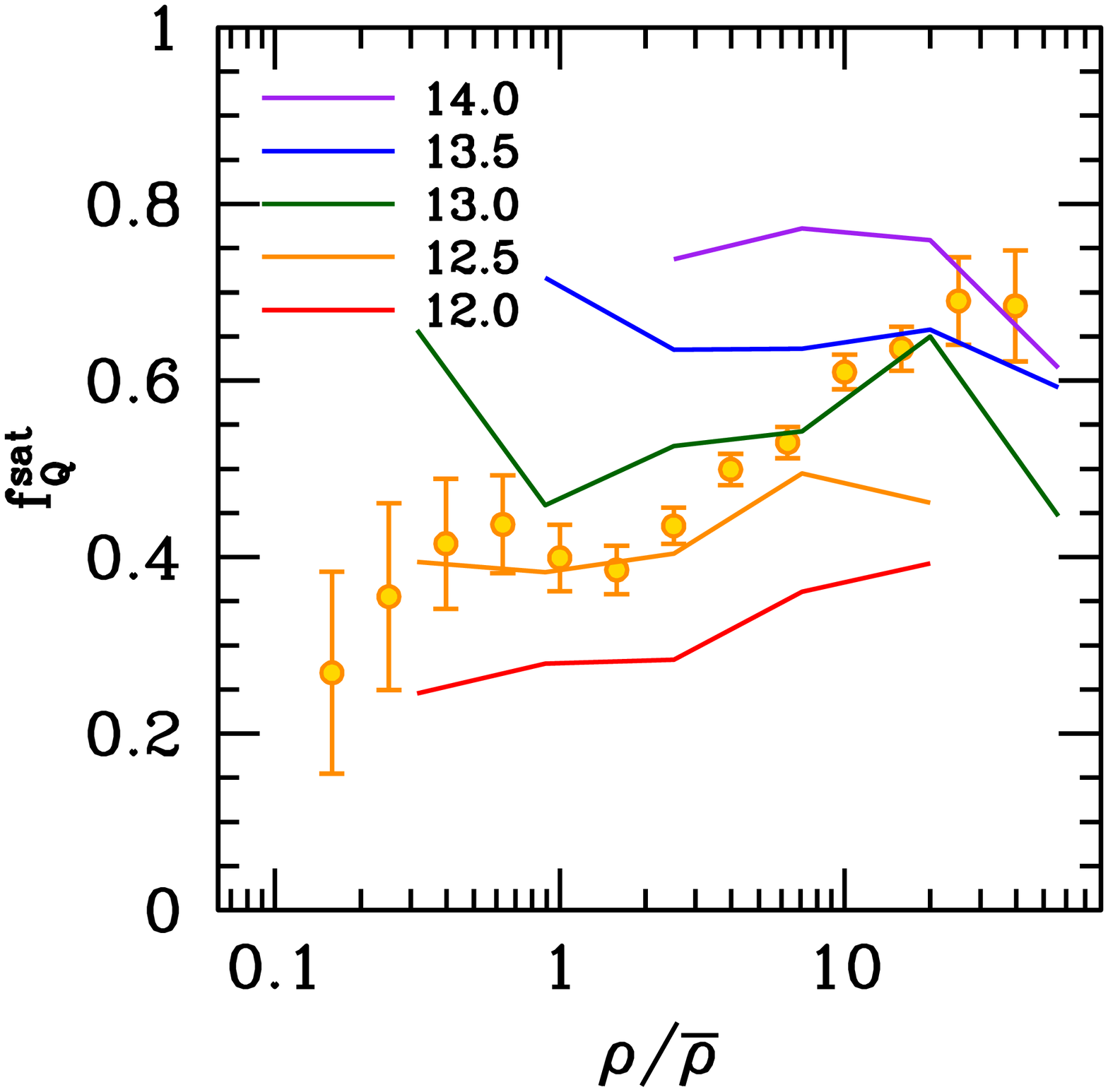,width=8.0cm}}
\caption{ \label{fq_den_sats} Quenched fraction of satellites,
  $\fqsat$, as a function of 10 \hmpc\ overdensity, $\rhob$, for
  satellites with $M_r = [-19,-20]$.  Filled circles show the
  measurements from Fig.~\ref{fq_den_decomp}. Solid curves show
  $\fqsat$ in bins of $\log\mhost$.}
\end{figure}

\begin{figure}
\vspace{-1.0cm} 
\centerline{\psfig{figure=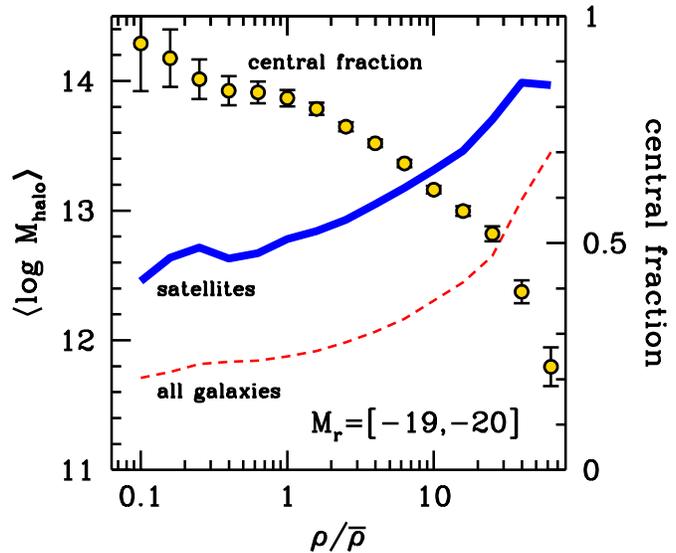,width=9.0cm}}
\caption{ \label{fcen_hostmass} Solid and dashed curves indicate the
  mean $\log\mhost$ as a function of 10 \hmpc\ 
  overdensity, $\rhob$, for satellite galaxies and all galaxies in the
  $M_r=[-19,-20]$ bin, respectively. Here,
  $\langle \log\mhost \rangle$ is a galaxy-number weighted mean.
  Unlike central galaxies, the typical halo mass probed by satellite 
  galaxies increases with density.
  The filled circles indicate the fraction of all galaxies that are
  central galaxies as a function of $\rhob$, as given by the right-hand 
  $y$-axis.  }
\end{figure}

\subsection{Properties of Satellite Galaxies at Fixed Halo Mass}

For our fiducial results we examine a bin in magnitude of
$M_r=[-19,-20]$. These galaxies are faint enough that they span a
wide range of host halo masses (as satellite galaxies) but bright
enough that a volume-limited sample contains sufficient
statistics for fine binning in halo mass. By looking at galaxies in a
relatively narrow range of luminosities, we are restricting our
analysis to galaxies that are in a narrow range of $\msub$ (for
satellite galaxies) and $\mhost$ (for central galaxies).

Fig.~\ref{fquench_hostmass} shows the fraction of quenched satellites,
$\fqsat$, as a function of $\mhost$ for bins in $M_r$.  The slope of
$\fqsat$ is independent of galaxy magnitude, but the amplitude
increases monotonically with $M_r$.  These results indicate that
accretion of a galaxy onto a larger halo contributes significantly to
the buildup of quenched, passive galaxies, in agreement with previous
works (e.g., \citealt{vdb_etal:08, tinker_wetzel:10}).

The satellite trends in Fig.~\ref{fquench_hostmass} are broadly
consistent with previous results on galaxy mass and halo mass
dependence.  Using the maxBCG cluster sample, \cite{hansen_etal:09}
found that the red fraction of galaxies increases with cluster
richness, though their red fraction is somewhat higher than our
quenched fraction due to the use of color rather than a
dust-insensitive diagnostic.  Other works using more direct star
formation rate indicators have found more similar trends
\citep{weinmann_etal:06b, kimm_etal:09, weinmann_etal:10,
  vonderlinden_etal:10}.  We will examine trends of satellite
quenching in much more detail in Papers II and III. For now, these
results will be important when interpreting the correlations with
density in Fig.~\ref{galprop_density}.

\subsection{Dissecting the Correlations with Density}

Fig.~\ref{fq_den_4win} shows the $\fq$-$\rhonorm$ relation, broken
into central and satellite galaxies, for four magnitude bins.  For
$L\lesssim L_\ast$ galaxies, the quenched fraction of central galaxies
is independent of large-scale density, spanning the entire range of
environments from the deepest voids $(\rho/\rhobar\sim 0.1)$ to
cluster infall regions $(\rho/\rhobar\gtrsim 10)$. The entire
correlation with environment is driven by the satellite galaxies. The
results are consistent with the scenario in which all environmental
correlations are due to the change in the halo mass function. However,
for the $M_r=[-20.5,-21]$ magnitude bin there is a clear dependence of
$\fqcen$ with density, increasing from $\fq=0.55$ to $0.65$ over two
decades in density. We will discuss the brighter galaxies
subsequently, focusing now on understanding the $\fq-\rhonorm$
correlation for our fiducial sample.

The left panel in Fig.~\ref{fq_den_decomp} shows the
$\fq$-$\rho/\rhobar$ relation for $M_r=[-19,-20]$ galaxies taken from
Fig.~\ref{fq_den_4win}b. The dotted curves indicate the raw quenched
fractions, uncorrected for satellite-central mislabeling.  Because
mislabeling only increases $\fqcen$ and decreases $\fqsat$, these
curves can be considered upper and lower limits on these two
quantities, respectively. The right panel shows the same breakdown for
groups defined by stellar mass rather than luminosity. In both panels,
the dependence of $\fq$ on environment is caused entirely satellite
galaxies; $\fqsat$ rises rapidly when $\rhob>1$.

Understanding this satellite dependence on $\rhonorm$ requires a
decomposition into host halo mass.  As was demonstrated in
Fig.~\ref{fquench_hostmass}, the quenched fraction of satellite
galaxies at fixed luminosity is a strong function of
$\mhost$. Fig.~\ref{fq_den_sats} breaks $\fqsat$ into bins of fixed
$\mhost$. When controlling for halo mass, the quenched fraction of
satellites also is independent of environment within the statistics of
the sample.  Thus, the observed correlation between $\fq$ and
$\rhonorm$ simply arises from the change in the halo mass function
with environment.

Fig.~\ref{fcen_hostmass} shows the relative number of centrals and
satellites as a function of large-scale environment. Unsurprisingly,
central galaxies dominate the statistics in the voids. As
$\rho/\rhobar$ increases, the central fraction monotonically
decreases, dropping under 50\% at $\rhonorm>20$. Because we are
looking at a bin in galaxy magnitude, the halo mass probed by centrals
is constant with density at $\log M\sim 11.7$, but the typical halo
mass probed by satellite galaxies monotonically increases with
density. At 10 times the mean galaxy density, $\langle \log\mhost
\rangle\sim 13$ for satellite galaxies.

The change in $\langle \log\mhost \rangle$ with large-scale density is
a well-understood phenomenon: more massive dark matter halos are more
likely form in higher overdensities (e.g., \citealt{bond_etal:91,
  sheth_tormen:02}).  Fig.~\ref{fquench_hostmass} demonstrates that
$\fq$ depends on $\mhost$, thus as $\rho$ increases the quenched
fraction of satellites increases as a result of the changing halo mass
function.

\subsection{Bright Central Galaxies}

\begin{figure}
\centerline{\psfig{figure=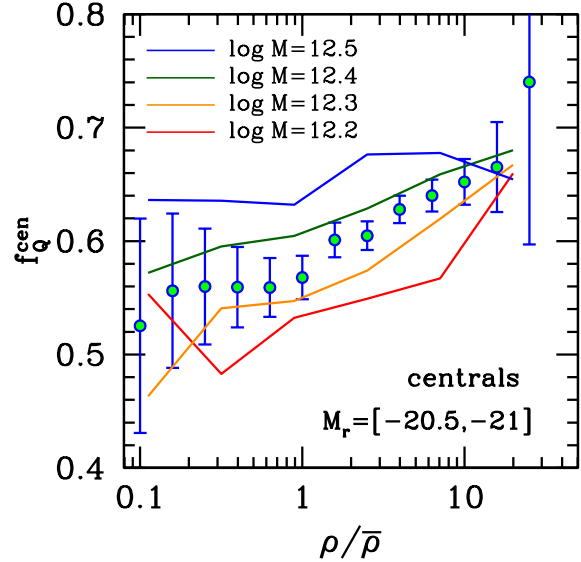,width=8.0cm}}
\caption{ \label{color_density_centrals} Quenched fraction of bright
  central galaxies vs. 10 \hmpc\ overdensity. The circles represent
  $\fqcen$ for all central galaxies in this magnitude bin, while the
  curves show these galaxies split into different halo masses,
  demonstrating that the increase in quenched fraction with density
  also persists for fixed halo mass.}
\end{figure}

As demonstrated in Fig.~\ref{fq_den_4win}, the quenched fraction of
$L>\lstar$ central galaxies does exhibit a weak but significant
correlation with density. Using our Poisson error bars, a line with
slope $d\fqsat/d\log \rho= 0.08$ yields a $\Delta \chi^2$ of 25
relative to the best-fit constant value. Due to scatter between $M_r$
and $\mhalo$, central galaxies in this magnitude bin span a range in
halo mass. Thus we must determine if this correlation is due to an
intrinsic relationship between $\fq$ and density or a change in the
typical halo mass probed in low and high
densities. Fig.~\ref{color_density_centrals} shows the results for the
$M_r=[-20.5,-21]$ central galaxies broken into bins of fixed halo
mass. For each halo mass bin, the quenched fraction shows a
correlation with density comparable to that of the overall magnitude
bin, which demonstrates that more massive galaxies exhibit a true
correlation with environment at fixed halo mass.

\begin{figure*}
\centerline{\psfig{figure=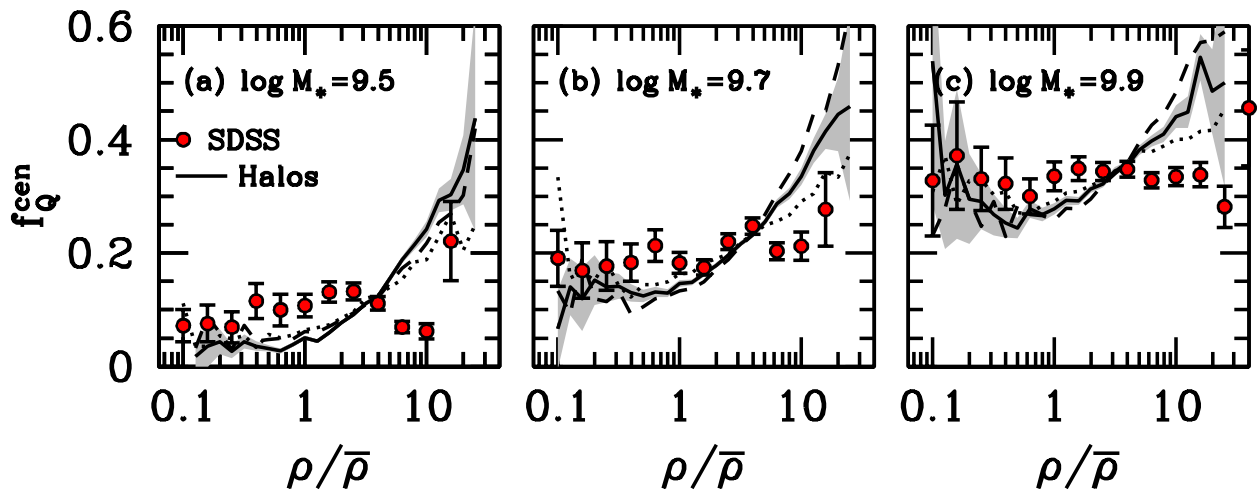,width=18.0cm}}
\vspace{-10.5cm}
\caption{ \label{mdot_delta3} Central quenched fraction, $\fqcen$ vs.
  10 \hmpc\ overdensity, $\rhob$, in three bins of stellar
  mass. Filled circles indicate the measurements from the group
  catalog.  The mean logarithmic halo mass for each stellar mass bin
  is 11.4, 11.5 and 11.6 for panels (a), (b), and (c), respectively.
  Curves indicate the `old' fraction of dark matter halos in the
  corresponding mass ranges, where `old halos are defined as those
  with the lowest growth since a given redshift, and the growth
  threshold is chosen to match the corresponding fraction of quenched
  central galaxies.  The solid curves show halos with lowest growth
  from $z=0.5$ to $z=0$, while the dotted and dashed curves are from
  $z=1.0$ and $z=0.3$, respectively. For halos, $\rhob$ is measured in
  the same manner as the data, using galaxy density on the 10 \hmpc\
  scale. Galaxies are placed into halos and subhalos using
  Eq. (\ref{e.sham}). All errors are Poisson.  Galaxies with the
  lowest growth rates do not simply reside in halos with lowest growth
  rates. }
\end{figure*}

\section{Confronting Theory with Observations}

\subsection{Quenched Galaxy Fractions vs.~Old Halo Fractions}

Although a full model of galaxy formation requires incorporation of
many unsolved physical processes, it is possible to test the ansatz
that the oldest (or slowest growing) galaxies reside in the oldest
halos. In the context of the observational metric used here,
`old' galaxies have ceased growing in mass through star
formation\footnote{Low-mass, quenched galaxies may increase their mass
  through merging, but significant mass growth through mergers is
  unlikely for sub-$\lstar$ galaxies (\citealt{maller:08,
    conroy_wechsler:09}) and major mergers would more likely involve a
  gas-rich galaxy, leading to significant star formation.}, while `old'
halos have accumulated most of their mass at early times and have
little present mass growth.

In this section we compare theoretical predictions to results from the
group catalog that uses stellar mass rather than luminosity to
estimate halo mass. We create predictions using halos in a
high-resolution $N$-body simulation described in Appendix C and Paper
III. This simulation has sufficient mass and spatial resolution to
track the evolution of halos and subhalos down to $10^{11}$ \hmsol. We
populate the halos and subhalos of this simulation with galaxies using
the abundance matching technique described in \S 2, assigning galaxy
luminosities with the \cite{blanton_etal:03} luminosity function. This
allows us to calculate $\rhonorm$ for objects in the simulation in the
same manner as our DR7 results\footnote{When measuring $\rho$ for
  galaxies in the stellar mass limited group sample, we use the full
  volume-limited galaxy sample (to the magnitude limit of the volume)
  in order to maximize the statistics with which we measure density.}.

Fig.~\ref{mdot_delta3} shows $\fq$-$\rhonorm$ for three bins in galaxy
mass (all less massive than the knee in the stellar mass function). In
each stellar mass bin, the quenched fraction is independent of
environment, consistent with the results from from \S 4. From the
group catalog, these galaxies live in halos of mass $\log\mhost=11.4$,
$11.5$, and $11.6$. We bin the halos in the simulation to match the
halo masses probed in the stellar mass samples in
Fig.~\ref{mdot_delta3}. In each bin, the halos are divided into `old'
and `young' using the fractional growth rate of each halo. In each
panel, we plot the old fraction of halos, $\fold$, as a function of
$\rhonorm$, normalized such that the overall old fraction of halos in
each mass bin is the same as $\fqcen$ of the galaxies. In each panel,
the $\fold$ depends strongly on $\rhonorm$ such that older halos are
preferentially found in high-density environments. This is consistent
with earlier numerical results on assembly bias in which old, low-mass
halos are more highly clustered than young halos of the same mass (see
the references in \S 1).  The different curves in each panel represent
different redshift baselines over which to calculate the halo growth
rate, defined\footnote{Note that the halo mass used, $M(z)$, is the
  {\it maximum} mass the halo has achieved up to redshift $z$. The
  maximal mass should correlate closer to galaxy stellar mass than
  instantaneous mass (\citealt{wetzel_white:10}). Halos that lose
  mass from redshift $z$ to zero would then have growth rates of 0. The
  rank-ordering of halos by growth rate is insensitive to choice
  of instantaneous versus maximum mass.} as $1-M(z)/M(0)$, where
$z=0.3$, 0.5, and 1.0.  Using a shorter redshift baselines leads to a
somewhat stronger environmental dependence.  But regardless of the
definition of halo growth rate, the trend for the halos is not matched
by the galaxies. This implies that halo formation history is not
correlated with galaxy formation history; at fixed stellar mass, old,
quenched galaxies and young, star-forming galaxies live in halos drawn
effectively randomly from the distribution of halo growth rates.

\begin{figure}
\centerline{\psfig{figure=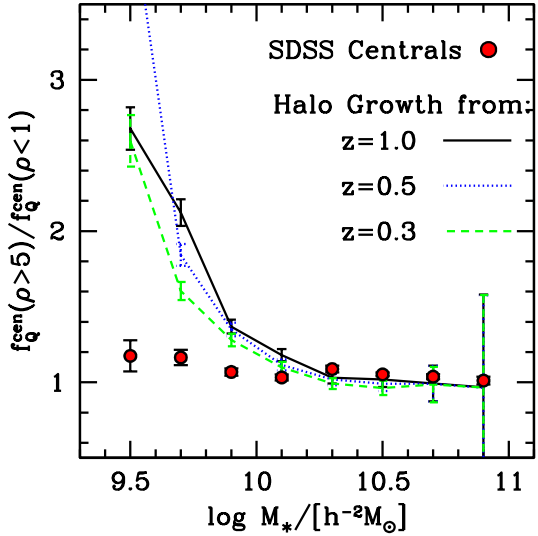,width=8.0cm}}
\caption{ \label{mdot_slopes} The ratio of the central quenched fraction 
  between high- and low-density environments as a function of
  stellar mass. Filled circles show results from the group catalog, 
  while curves indicate results using the growth rates of
  corresponding dark matter halos as measured over different redshift 
  baselines. }
\end{figure}

To demonstrate this result more clearly, Fig.~\ref{mdot_slopes} shows
the ratio of the quenched fractions in high- and low-density regions
as a function of central galaxy stellar mass. The ratio for SDSS
centrals is near unity at all $\mgal$. For galaxies $\log\mgal\la 10$,
the results are inconsistent with halo growth being correlated with
galaxy growth.
 
\subsection{Halo Formation Times Versus the Redshift of Galaxy Quenching}

Rather than assuming a monotonic relationship between galaxy age and
halo age, we can also test whether halo formation times correlate with
inferred galaxy quenching times for central galaxies. We define
quenching time as the lookback time to the redshift at which a galaxy
last formed stars. This comparison is model-dependent, both on the
observational and theoretical sides, but we will use quantities and
parameters to maximize the validity of the comparison.

To model quenched galaxies with stellar population synthesis (SPS)
models, we assume a single burst of star formation. Using the SPS
models of \cite{conroy_etal:09} and \cite{conroy_gunn:10} we calculate
the time evolution of $\dn$. For these calculations, we assume an
initial mass function of \cite{chabrier:03} and solar metallicity. To
model actively star forming galaxies, we assume a star formation
history of $SFR \propto t\times \exp(-t/\tau)$ where $\tau=3.5$ Gyr,
consistent with the results of \cite{noeske_etal:07a} for the stellar
masses probed here (see Paper III for full details regarding the
determination of $\tau$). Once again, we assume solar metallicity in
order to calculate $\dn$ as a function time. With these models, we
create a lookup table to determine either the quenching timescale or
formation timescale for passive and active galaxies,
respectively. These timescales are model dependent, being especially
sensitive to the metallicity assumed. Thus, the quantity we compare to
theoretical predictions is the variation relative to the mean
timescale as a function of density. This quantity removes a
significant portion of the model dependence.

For dark matter halos, a common quantity used to measure the formation
time of a halo is $t_{1/2}$ the time since a halo accumulated half its
current mass. However, for this measure the peak in the age
distribution is $\sim 9.5$ Gyr ($z\sim 1.6$). Most low-mass quenched
galaxies have migrated to the red sequence at $z<1$
(\citealt{drory_etal:09}), demanding a measure of halo formation time
that probes more recent structure formation. To this end, we use
$t_{85}$ the time (in Gyr) since a halo reached 85\% of its maximal
mass. For this measure, the peak in the halo age distribution is $\sim
4$ Gyr ($z\sim 0.4$), but with a broad tail to higher redshift.

\begin{figure*}
\centerline{\psfig{figure=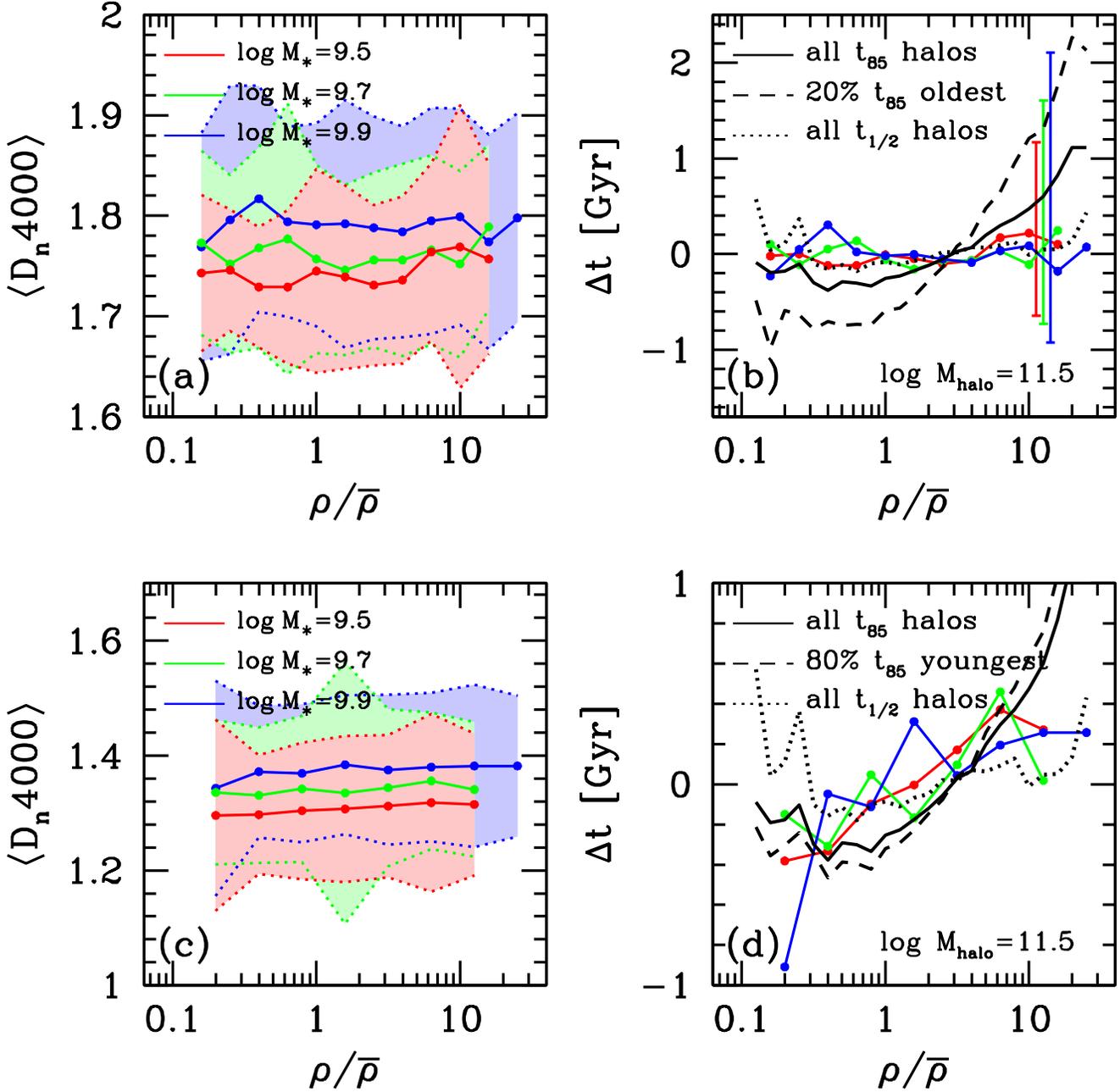,width=18.0cm}}
\caption{ \label{dn4k_age} Panel (a): $\dn$ values for quiescent
  central galaxies as a function of $\rhonorm$. The connected dots
  indicate the mean $\dn$ for different $\mgal$ bins. The shaded
  regions indicate the dispersion for each stellar mass bin. Panel
  (b): The change in the mean quenching timescale inferred from
  $\langle \dn\rangle$ in panel (a). Timescales are calculated from
  stellar population synthesis models described in the text. Here we
  plot $\Delta t$ rather than absolute age because it is less
  sensitive to model assumptions. Connected dots represent the same
  stellar mass bins as in (a). The errorbars indicate the dispersion
  in $\Delta t$ implied by the variance in $\dn$. Black curves
  represent results from dark matter halos: the solid black curve is
  the change in the mean formation time $t_{85}$ for halos of mass
  $10^{11.5}$ \hmsol (equivalent to the halo mass for $\mgal=10^{9.7}
  \hhmsol$ central galaxies). Dashed black curve shows $\Delta t$ for
  the oldest 20\% of halos at each density bin (equivalent to $\fq$
  for central galaxies with $\mgal=10^{9.7} \hhmsol$). The dotted
  curve shows $\Delta t$ for halos based on $t_{1/2}$. Panel (c): same
  as (a), but now for star-forming central galaxies (defined as
  $\dn<1.6$). Panel (d): Same as (b) but again for star-forming
  galaxies. In this panel, the dashed curve indicates $\Delta t$ for
  the 80\% youngest halos at each density (corresponding to the
  fraction of star-forming central galaxies). Note the change in
  $y$-axis range.}
\end{figure*}

Fig.~\ref{dn4k_age}a shows the mean $\dn$ values as a function of
$\rhonorm$ for low-mass quenched central galaxies ($\dn>1.6$). The
mean $\dn$ does not vary with density, implying that the quenching
time is constant with density as well. The mean quenching timescale is
constant to within $\sim 100$ Myr across all
densities. Fig.~\ref{dn4k_age}b compares these results to the
formation times of dark matter halos. The solid curve is the change in
$t_{85}$ relative to the mean for all halos with $\log \mhalo=11.5\pm
0.15$ (equivalent to the host halo masses for galaxies with
$\mgal=10^{9.7} \hhmsol$). In this comparison, the key assumption is
not that the oldest halos contain quenched galaxies, but rather that
the quenching time correlates with halo formation epoch. Under this
assumption, the mean quenching timescale should vary by roughly 1.3
Gyr with density. If we combine these two assumptions we obtain the
dashed curve---here, we isolate the oldest 20\% of galaxies in each
bin in $\rho$ and determine $\Delta t_{85}$ for only these halos. In
this scenario, the mean formation time varies by nearly 3 Gyr from low
to high densities.

The dotted curve plots $\Delta t_{1/2}$ rather than using $t_{85}$,
which shows no correlation with density. However, as we argue above,
using a timescale that probes structure at $z>1$ is unlikely to have
any correlation with the recent star formation and growth of galaxies,
in contrast to our chosen statistic of $t_{85}$. Recent halo growth
correlates strongly with current environment, but if one measures halo
growth so far in the past, these correlations become washed out unless
one probes the wings of the distribution.

Fig.~\ref{dn4k_age}c is parallel to \ref{dn4k_age}a, but now for
star-forming central galaxies ($\dn<1.6$). Once again, the mean $\dn$
is nearly independent of density. When converting to formation
timescale, however, a slight positive trend with $\rho$ is apparent;
the timescale varies by roughly 0.6 Gyr from low to high
densities. Results for dark matter halos show a stronger trend with
density, especially at $\rhonorm\ga 10$. Under the assumptions of our
stellar population synthesis models, star forming galaxies in high
densities are slightly older than their counterparts in
underdensities, but the amplitude of the effect is about half what one
would predict using the formation times of dark matter halos.

\subsection{Evolution of the Stellar-to-Halo Mass Ratio}

\begin{figure}
\centerline{\psfig{figure=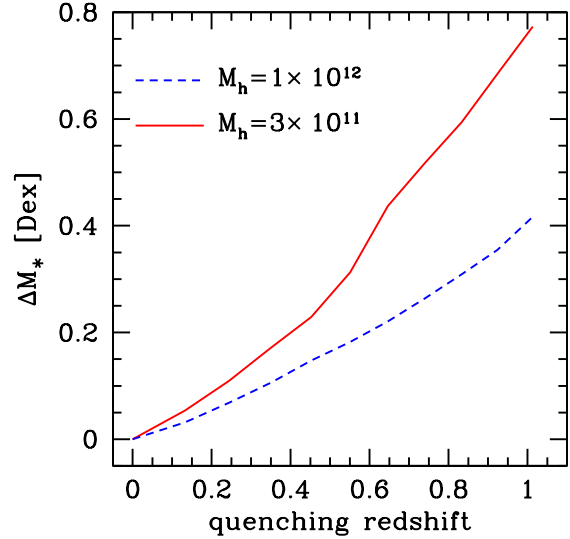,width=8.0cm}}
\caption{ \label{sham} The difference in logarithmic stellar mass 
  between the $z=0$ mean $\mgal$-$\mhalo$ relation and the relation for 
  galaxies that were quenched at redshift $z$, as a function of $z$. 
  This plot assumes that halo growth and galaxy growth
  are uncorrelated, thus when a galaxy is quenched the halo it resides
  in continues to grow at the mean rate up to $z=0$. Solid and dashed
  curves are for halos with $z=0$ masses of $3\times 10^{11}$ and 
  $10^{12}$ \hmsol, which host central galaxies of $\log\mgal=9.7$ and 10.2 
  \hmsol, respectively. }
\end{figure}

Figs.~\ref{mdot_delta3} and \ref{dn4k_age} imply that halo growth
rates are not correlated with galaxy growth rates for galaxies below
the knee in the stellar mass function---or, perhaps more
straightforwardly, that the mechanisms the quench galaxy star
formation are not correlated with dark matter growth. If halo growth
rate is indeed uncorrelated with galaxy growth rate, then the
$\mgal/\mhalo$ ratio for low-mass galaxies will evolve differently for
central quenched galaxies as compared with those that remain active.
For quenched galaxies, $\mgal$ is essentially fixed while $\mhalo$
continues to evolve, while active galaxies continue to grow at a much
higher fractional rate than their halos (see, e.g., Fig.~4 in
\citealt{conroy_wechsler:09}). Assuming that a halo is on the mean
$\mgal/\mhalo$ relation at the time its star formation is quenched,
Fig.~\ref{sham} shows how far off the mean $\mgal/\mhalo$ relation at
$z=0$ that halo would be as a function of the redshift of
quenching. The details of this calculation are as follows: first, we
set up the $\mgal/\mhalo$ ratio as a function of halo mass through
Eq. (\ref{e.sham}), using the halo plus subhalo mass function in
Eq. (\ref{e.mf_smf}).  For stellar mass, we use the
\citet{li_white:09} stellar mass function at $z=0$, and we linearly
interpolate to higher redshift stellar mass functions as given in
\cite{marchesini_etal:07}.  We consider two different $z=0$ halo mass
bins: $\mhalo=10^{12}$ \hmsol\ and $10^{11.5}$ \hmsol. Using our
$N$-body simulation, we track these halos back in time and determine
their masses as a function of redshift, assigning them galaxy masses
that lie on the $\mgal/\mhalo$ relation at each redshift. Assuming a
galaxy is quenched at redshift $z$, we calculate the difference
between the stellar mass at $z$ and the mean relation at $z=0$ for
each halo mass bin. The earlier a galaxy is quenched, the larger the
difference with respect to the the typical galaxy mass for that halo
mass. The currently best-estimated lognormal scatter in the
$\mgal$-$\mhalo$ relation at $z\sim0$ is $0.15-0.2$ dex
(\citealt{more_etal:10, leauthaud_etal:11b}), so the offset in
Fig.~\ref{sham} becomes larger than the scatter for centrals quenching
prior to $z\approx0.3$.

\section{Discussion}

\subsection{The Environmental Dependence of Galaxy Properties}

In this paper we have demonstrated that the properties of galaxies at
fixed halo mass are independent of density measured on scales larger
than the halo radius, and that the dependence of the quenched fraction
of galaxies on large-scale environment can be explained by the
variations of the halo mass function with environment and by
disentangling the relative contributions of central and satellite
galaxies.  Our results extends earlier works that demonstrated that
galaxy properties correlate weakly with large-scale environment when
small-scale environment is fixed on a 1 Mpc scale
(\citealt{kauffmann_etal:04,blanton_etal:06a}).  However, we have
argued that, because different galaxies live in different halos, it is
not possible (or physically motivated) to find a single smoothing
scale that fully determines galaxy properties at all masses.  For rich
groups and clusters, our results are in good agreement with
\cite{blanton_berlind:07}, who found that the the blue fraction of
galaxies is independent of density at fixed group luminosity, though
we do see evidence for some change with density for the brightest
central galaxies.  These results are in good agreement with the
results of \cite{peng_etal:11}, who also find that the environmental
dependence of the red fraction of galaxies is driven primarily by
satellite galaxies and not by centrals. In detail, the observations
diverge from this scenario for quenched bright central galaxies and in
the mean stellar ages of central galaxies on the star-forming
sequence. But our results support the notion that halo mass is the
dominant quantity governing the properties of the galaxies contained
within them.

\subsection{Halo Assembly Bias and the $\fq$-Density Relation}

As discussed in \S 1, the formation histories of dark matter halos
depend on environment. For low-mass halos, old halos form in
high-density environments. For high-mass halos, old halos form in
low-density environments, although there is some debate on the
amplitude of this correlation. If halo and galaxy formation histories
are correlated, then our measurements are inconsistent with both of
these theoretical results. Figs.~\ref{fq_den_decomp} and
\ref{mdot_delta3} clearly demonstrate a lack of correlation between
$\fq$ and $\rhonorm$ for low-mass galaxies. Fig.~\ref{dn4k_age}
demonstrates that the mean stellar ages of quenched central galaxies
also is constant with $\rhonorm$. For galaxies below the knee is the
stellar mass function, the ansatz that old galaxies live in old halos
is not supported by the data.

However, the quenched fraction of central galaxies in $\mhost\sim
10^{12.3}$ \hmsol\ halos show a positive correlation with
environment. This is the mass scale at which halos in simulations show
little-to-no correlation between age and $\rhonorm$. One possibility
is that the environmental dependence for high-mass central galaxies
arises from those that have passed through the virial radius of a more
massive halo (and therefore reside higher density environments) but
are counted as centrals at the present time
\citep{gill_etal:05,ludlow_etal:09,wang_etal:09}.  However, these
ejected satellites are expected to become less common at increasing
galaxy mass, so the effect is opposite to the trend we see here (we
will discuss these processes in more detail in Papers II and III).
Another possibility is that our assumption that the central galaxy is
the most massive galaxy is becoming less valid at higher galaxy mass,
as suggested by \citet{skibba_etal:11}, and we are confusing high-mass
centrals and satellites.  However, in order to masquerade as the
signal in Fig.~\ref{color_density_centrals}, this effect would have to
have strong environmental dependence, which remains
unclear. Additionally, at higher halo masses, central and satellite
galaxies have increasingly similar quenched fractions. Thus, the
stellar mass dependences of \cite{skibba_etal:11} and central/satellite
quenched fractions should cancel out at some level.

Our results have mixed agreement with those of
\cite{wang_etal:08_assembly_bias}, who used a similar group-finding
algorithm. They found that galaxy groups with redder overall colors
have higher clustering than groups of the same mass but with bluer
galaxies, in agreement with our results.  However, they also found
that the effect is stronger in less massive groups, opposite to the
mass dependence we see here. Wang et.~al.~use color as a proxy for age
as opposed to (dust-insensitive) $\dn$, though it is unclear why using
color would lead to a stronger signal.

\subsection{The Stellar to Halo Mass Ratios of Low-Mass Galaxies}

Galaxies that are star forming and galaxies that are quenched form in
halos that growth the same average rate. This implies that the stellar
mass to halo mass ratio, $\mgal/\mhalo$ will evolve differently for
quenched and star-forming galaxies. Fig.~\ref{sham} demonstrates a
$\mhalo=10^{11.5}$ \hmsol\ halo that contains a central galaxy
quenched at $z=0.6$ will have a $\mgal/\mhalo$ ratio lower by a factor
of 2.5 at $z=0$ relative to the mean mean for that halo mass. However,
weak lensing and satellite kinematical measurements of the halos
masses of low-mass galaxies find that the halo masses are the same
within the errors, regardless of color selection
(\citealt{mandelbaum_etal:06_gals, more_etal:10}). This is in conflict
with the conclusion that galaxy growth and halo growth are
uncorrelated. There are several possible resolutions to this
discrepancy.

\begin{enumerate}
\item{The observations of halo masses of low-mass central galaxies are
    biased due the use of color rather than $\dn$ for sample
    segregation. The low-mass red samples will contain many
    intrinsically star-forming galaxies, thus bringing the mean halo
    masses closer together. Lack of significant numbers also increase
    observational errors and low masses.}
\item{Low-mass galaxies on the red sequence were quenched recently,
    and thus have not had enough time to create a significant
    difference in the $\mgal/\mhalo$ ratio. There is a clear red
    sequence at $z=1$, but the fraction of quenched central galaxies
    at $\mgal\la 10^{10}$ is not known.}
\item{Sub-$\lstar$ central galaxies do not follow a one-way path to
    the red sequence; galaxies migrate back and forth from the blue
    cloud to the red sequence before permanently becoming
    quenched. Some blue galaxies at $z=0$ may have been temporarily
    quenched previously, while some $z=1$ red galaxies may still grow
    through star formation later.  This would minimize the difference
    between the halo masses of red and blue central galaxies at fixed
    $\mgal$.}
\end{enumerate}

Galaxies that are quenched, for example, by AGN heating or major
galaxy mergers (e.g., \citealt{croton_etal:06a, bower_etal:06,
  hopkins_etal:08b, somerville_etal:08}) could still accrete gas or
merge with gas-rich objects at some later time, replenishing their
fuel supply and moving them back into the star-forming sequence. In
hydrodynamical simulations of galaxy formation, low-mass halos
($\mhalo\lesssim 10^{12}$ \hmsol) accrete their gas in a cold state
(\citealt{keres_etal:05, dekel_birnboim:06, keres_etal:09,
  brooks_etal:09}). Thus previous merging or AGN activity that might
temporarily quench a galaxy will not prevent any new gas from
migrating to the center of the potential well within a dynamical time,
even if the stellar population is old. However, none of these
solutions are mutually exclusive, and it is possible that a all three
cause the measurements presented here.

\subsection{What Quenches Low-mass Central Galaxies?}

The fraction of quenched central galaxies in the deepest voids is the
same as that in regions around clusters.  Thus the mechanism by which
low-mass central galaxies stop their star-formation must be
environmentally independent. \cite{wang_etal:09_dwarfs} note that many
red dwarf galaxies are isolated, but propose that the majority of
quenched central galaxies of this type are ejected satellites,
low-mass halos that have passed through a group or cluster. The
uniformity of $\fqcen$ for low-mass galaxies argues strongly against
such a model due to the strong environmental dependence that would
result from such a model. We cannot rule out two mechanisms that
operate primarily at low and high density, respectively, that add to a
constant $\fqcen$, but Occam's razor argues against such a scenario.

We find that the Sersic indices of these galaxies does not vary with
density, with $\langle n_s\rangle\approx 3.5$. This implies that the
mechanism that quenches low-mass centrals also induces morphological
transformation. It also strengthens the claim that the process that
halts star formation in centrals at high densities also operates at
low densities.

Using semi-analytic models, \cite{croton_farrar:08} find a small
population of central galaxies quenched by AGN heating, but this
process can only effect high-mass halos of $\mhalo>10^{12.5}$ \hmsol,
roughly an order or magnitude more massive than the host halos of the
galaxies probed here, and even more massive than the dwarf galaxies in
\cite{wang_etal:09_dwarfs}.

Halo mergers vary only weakly with large-scale dark matter
density (\citealt{fakhouri_ma:09}), implying that the galaxy merger rate
would also show no environmental dependence. Major mergers are also
likely to cause a transition from disk-dominated to
spheroidal-dominated morphology (see, e.g.,
\citealt{hopkins_etal:08b}). But it remains unclear whether there are
enough mergers to fully account for these objects. We leave this to a
future study (Wetzel et.~al., in preparation).

Further analysis of the morphologies and other properties of these
galaxies, as well as their redshift evolution, will shed light on the
mechanisms by which quenching occurs in these systems, and whether
or not quenching is transient.

\vspace{1.0cm}

\noindent The authors would like to thank Frank van den Bosch for
supplying the Yang et.~al. group catalog, as well as useful
discussions on the results presented in this paper.

\appendix
\section{Real vs. Redshift-Space Density Measures}

\begin{figure}
\centerline{\psfig{figure=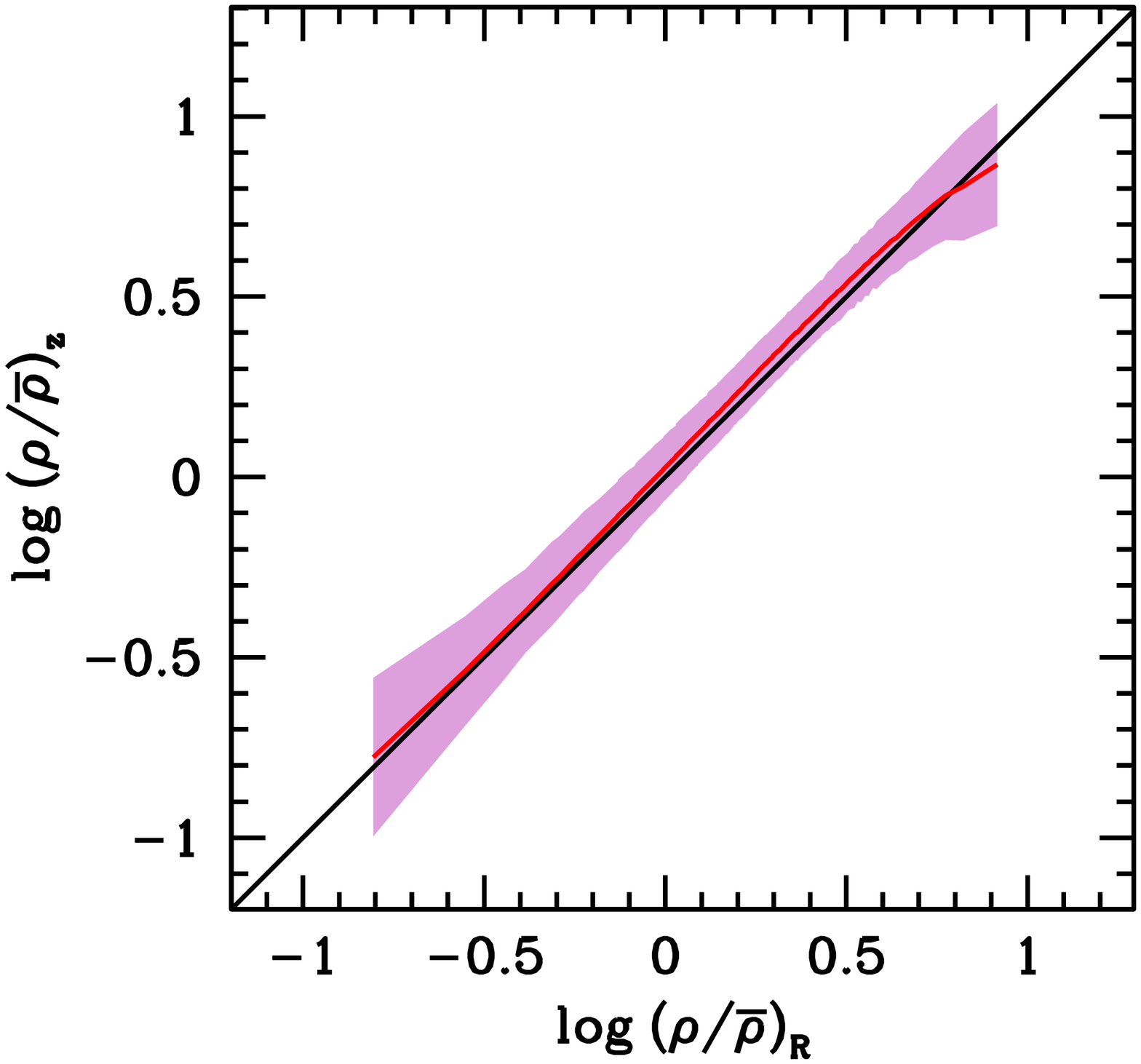,width=8.0cm}}
\caption{ \label{density_measures} Comparison between galaxy
  density in redshift space and real space as measured in a sphere of 
	radius 10 \hmpc. Results are presented for mock galaxies within a $720^3$
  $\hcubed$ simulation volume. The solid red curve shows the mean
  relation, and the 
  shaded region shows the dispersion about the mean. The solid black
  curve shows where the two measures are equal.  }
\end{figure}

Here we test the correlation between between large-scale galaxy
density measured in real and redshift space. We use the the numerical
simulation of \cite{wetzel_white:10} in which dark matter (sub)halos
are matched with galaxy luminosity by the abundance-matching method
described in \S 2, and the subhalos are populated into a larger volume
simulation as described in \citet{tinker_wetzel:10}.  The volume of
the simulation if $720^3$ $($\hmpc $)^3$, larger than the $M_r<-20$
volume limited sample from Table 1. For each galaxy in the simulation,
we determine the galaxy density in 10 \hmpc\ spheres in both real
space and redshift space, assuming the distant observer approximation
thus using the $z$-axis of the simulation as the line of sight.
Fig.~\ref{density_measures} shows the correlation and scatter between
the two. At this scale, the mean redshift-space density faithfully
tracks the real-space density. More importantly, the scatter is
generally $\lesssim 0.1$ dex within the extremes of the density field.

\section{Group Finding Algorithm}

Here we describe our galaxy group finding algorithm in more detail.
We define dark matter such that the mean density $\Delta$
interior to a radius $R_\Delta$ is 200 times the universal
matter density $\bar{\rho}=\om\times\rhoc$. Here and throughout this paper,
$R_\Delta$ is in comoving units. Thus the mass and radius of a halo
are related by

\begin{equation}
\label{e.halodef}
M_\Delta = \frac{4}{3}\pi R^3_\Delta\Delta\bar{\rho}.
\end{equation}

\noindent In comoving units, $\bar{\rho}$ is a constant, thus the
comoving radius of a halo is independent of redshift at fixed
mass. Assuming the virial theorem, the velocity dispersion of the dark
matter within a halo is

\begin{equation}
\label{e.veldisp}
\sigma_v^2 = \frac{GM_\Delta}{2R_\Delta}(1+z),
\end{equation}

\noindent where $G=4.304\times 10^{-9}$ in units of
(km/s)$^2$Mpc/M$_\odot$, and the factor of $(1+z)$ is to convert the
comoving radius to physics units.

Dark matter halos follow a `universal' density profile described in
\cite{navarro_etal:97} (hereafter NFW), defined by a scale radius
$r_s$ and a concentration parameter $c_\Delta=R_\Delta/r_s$. The 2-D projected
density profile of the NFW profile is given by

\begin{equation}
\Sigma(R)= 2~r_s~\bar{\delta}~\bar{\rho}_{\rm gal}~{f(R/r_s)}\,,
\end{equation}
with 
\begin{equation}
\label{fx}
f(x) = \left\{
\begin{array}{lll}
\frac{1}{x^{2}-1}\left(1-\frac{{\ln
{\frac{1+\sqrt{1-x^2}}{x}}}}{\sqrt{1-x^{2}}}\right)   &  \mbox{if   $x<1$}  \\
\frac{1}{3}   &   \mbox{if   $x=1$}   \\   
\frac{1}{x^{2}-1}\left(1-\frac{{\rm
      atan}\sqrt{x^2-1}}{\sqrt{x^{2}-1}}\right) & \mbox{if $x>1$}
\end{array} \right.\,,
\end{equation}
and 
\begin{equation}
\bar{\delta} = {200 \over 3} {c_{200}^3 \over {\rm ln}(1 + c_{200}) -
c_{200}/(1+c_{200})},
\end{equation}

\noindent where $\bar{\rho}_{\rm gal}$ is the mean galaxy density in
the volume-limited sample.  Thus, in constructing galaxy groups, we
assume that halos defined by a mean overdensity with respect to dark
matter of 200 will also appear at overdensities in the galaxy density
field with $\Delta=200$.

We assume that the velocity dispersion of satellite galaxies follows a
Gaussian distribution with dispersion given in equation
(\ref{e.veldisp}). Thus the line-of-sight probability of a galaxy
being a member of a group is defined by

\begin{equation}
\label{e.pz}
p(\Delta v) = \frac{1}{\sqrt{2\pi}\sigma_v}\exp\left[\frac{-(\Delta v)^2}{2\sigma_v^2}\right],
\end{equation}

\noindent where $\Delta v\equiv (z_{\rm gal} - z_{\rm group})/c$ and
$c$ is the speed of light, and the redshift of the group is defined as
that of the central galaxy.

All together, the number density contrast of a group of galaxies over
the background, as a function of both projected separation from the
central galaxy $R$, and line-of-sight velocity difference from the
central galaxy $\Delta v$, is expressed as

\begin{equation}
\label{e.psat}
P(\Delta v, R) = \frac{c}{H_0}\,\frac{\Sigma(R)}{\bar{\rho}_{\rm gal}}\,p(\Delta v).
\end{equation}

\noindent For a given galaxy, if $P$ is greater than a background
value $B$, then the galaxy is considered to be a member of the group.
We have calibrated $B=0.5$ via our simulation mock catalog to maximize
both completeness and purity.  We proceed through the galaxy catalog
in descending order of luminosity, and if a galaxy is classified as a
satellite, it cannot be a central galaxy (by definition) and thus
cannot have satellites of its own. In this algorithm, the brightest
(or most massive) galaxy in a group is the central galaxy by
definition.

Once galaxies have been assigned as centrals and satellites in dark
matter halos, we recalculate halo mass.  But we are now interested in
deconvolving the galaxy distribution into the occupation of galaxies
in host halos, we use the same abundance matching technique, but now
we use the {\it total} group luminosity and match that with the host
halo mass function (the \citealt{tinker_etal:08_mf} function). With
the new halo masses, we go back to the beginning and re-classify
galaxies as centrals and satellites, iterating the process until
convergence.

\begin{figure*}
\centerline{\psfig{figure=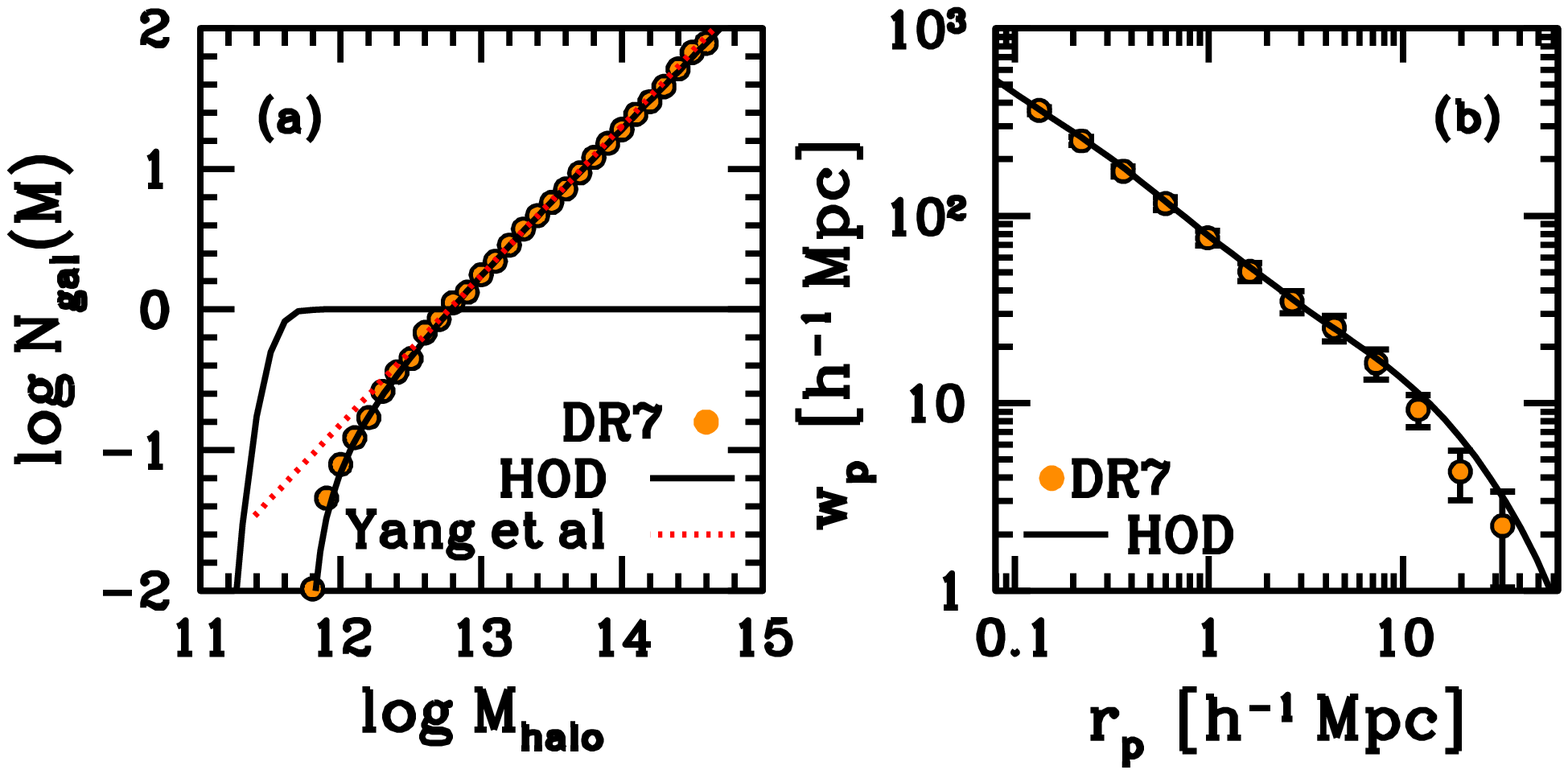,width=17.0cm}}
\vspace{-8.5cm}
\caption{ \label{wp_hod} Panel (a): Circles represent the halo
  occupation of satellite galaxies with $M_r<-19$ as a function of
  halo mass as measured from the group catalog. The solid curve shows
  a parametric HOD fitted to these data; the curve that asymptotes to $\log N_{\rm
    gal}=0$ represents central galaxies. The dotted line is the
  power-law fit to $N_{\rm gal}$ for the same sample of galaxies from
  \citealt{yang_etal:08} (their fit is restricted to $\mhalo\ga
  12.3$). Panel (b): Points with errors represent measurements of the
  projected two-point correlation function, $\wp$, from DR7 for
  $M_r<-19$ galaxies. The solid curve is the HOD prediction for $\wp$
  using the occupation function from panel (a).}
\end{figure*}

\section{Tests of the Group Finding Algorithm}

The group finder described above has been thoroughly tested by
\cite{yang_etal:05}. Because we are using a slightly different
implementation of the algorithm, and because we are probing
density-dependent statistics that have not been explored in 
previous papers, we perform two additional checks on our group
catalog.

\subsection{The Halo Occupation Distribution of the Group Catalog}

The aim of the group finding algorithm is to associate every galaxy in
the sample with a dark matter host halo, in the same framework as the
halo occupation distribution (HOD). For a fixed cosmology, a given HOD
maps uniquely onto a two-point correlation function. Thus a
consistency check on the results of the group finder is to compare the
observed correlation function of galaxies with that predicted by the
HOD measured from the group sample.

Fig.~\ref{wp_hod}a shows the the number of satellite galaxies with
$M_r<-19$ as a function of host halo mass. The dotted curve shows the
power-law fit to the same quantity from the \cite{yang_etal:08} SDSS
DR4 results. They do not include results from $\nsat<0.1$ host halos
in the fit.  The black curves shows a fit to both $\nsat$ and $\ncen$
using the HOD parameterization in
\cite{tinker_etal:08_voids}. Fig.~\ref{wp_hod}b shows the measured
projected two-point correlation function $\wp$ from DR7. Here we have
used the \cite{landy_szalay:93} estimator for $\wp$ and the random
catalogs provided in the VAGC to account for the survey angular
selection function. The black curve shows the predicted $\wp$ from the
HOD. We use the analytic model described in \cite{tinker_etal:05}. The
two correlation functions are in excellent agreement with one another,
demonstrating the consistency of the halo occupation of the groups.

\subsection{Recovery of Galaxy Properties on Mocks}

It is not possible to determine absolute membership of a group in
redshift space. Group members may have large velocities relative to
the group center, while non-members may be redshifted into the group
from coherent infall. Thus, some fraction of the group members are not
true members but are central galaxies that exist within a smaller
halo. Alternately, some galaxies classified as low-mass centrals are
actually members of a more massive group. (Satellites can also be
misclassified across groups, but this is comparatively rare.) The test
presented in Fig.~\ref{wp_hod} demonstrates that the {\it mean} number
of members is not biased, but because satellite galaxies have
different properties than central galaxies of the same magnitude, the
mean properties of observed group members will be biased. The true
quenched fraction of satellite galaxies is higher than the true
quenched fraction of centrals, thus the misclassification of
central/satellite galaxies can {\it only} have the effect of reducing
the observed $\fqsat$ and increasing $\fqcen$.

To test our group finding algorithm, we apply it to a mock galaxy
catalog from a 250 \hmpc\ $N$-body simulation with a particle mass of
$10^8$ \hmsol\ and force resolution of 2.5 \hkpc\
\citep[see][]{white_etal:10}, which robustly resolves the subhalos
that host galaxies we examine here. We describe the simulation,
subhalo finding, and mock catalog generation in detail in Paper
III. The cosmological parameters of this simulation are nearly
identical to those assumed in this paper. Galaxy luminosities are
assigned to halos and subhalos using the subhalo abundance matching
technique described in \S 2. As defined in \S 1, galaxies within
subhalos are labeled satellite galaxies, while all other galaxies are
central. The positions of the galaxies are transformed from real space
to redshift space using the distant observer approximation with the
$z$-axis of the box being the line of sight. 

\begin{figure*}
\begin{tabular}{cc}
\psfig{file=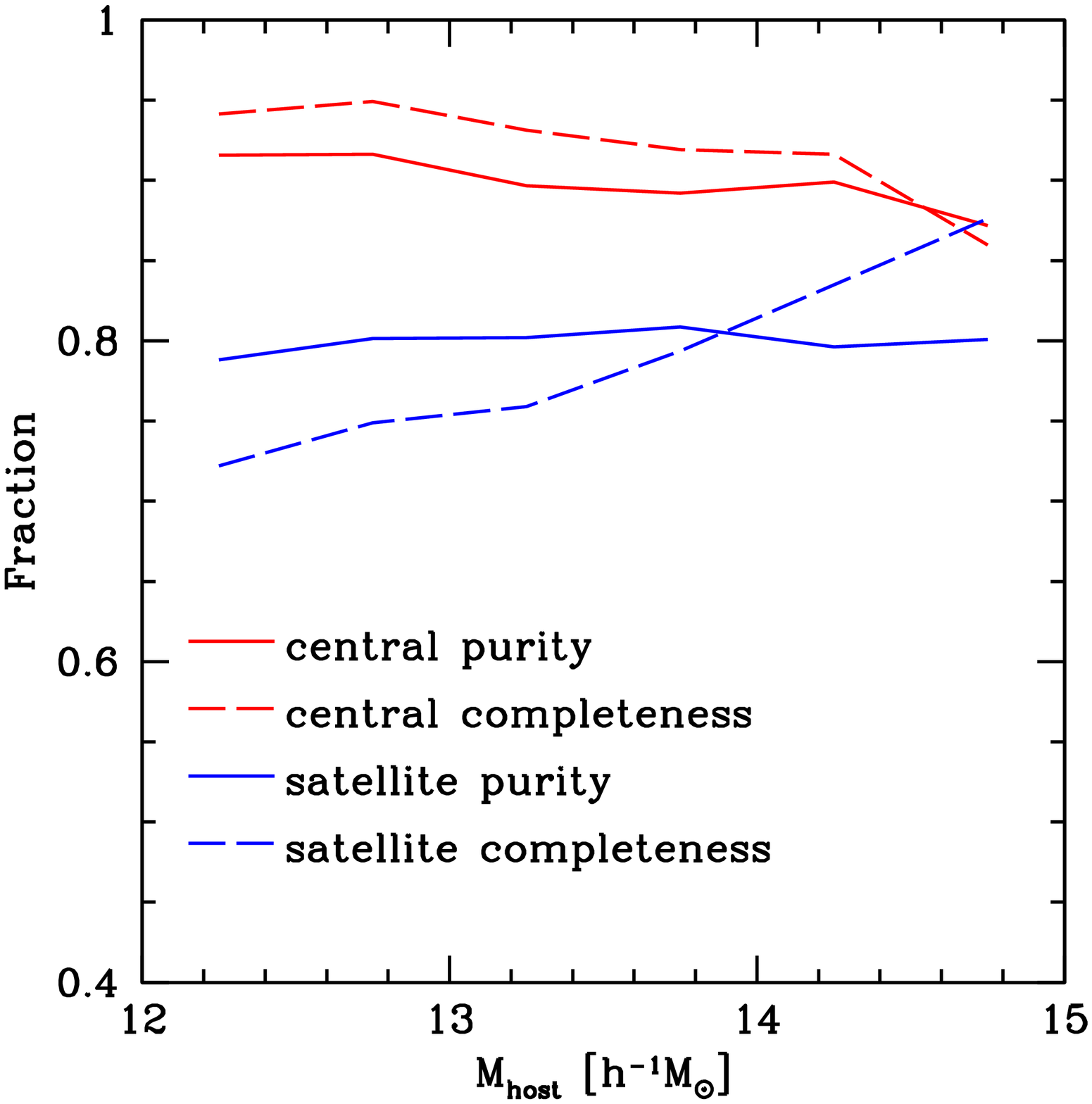,width=0.45\linewidth,clip=} &
\psfig{file=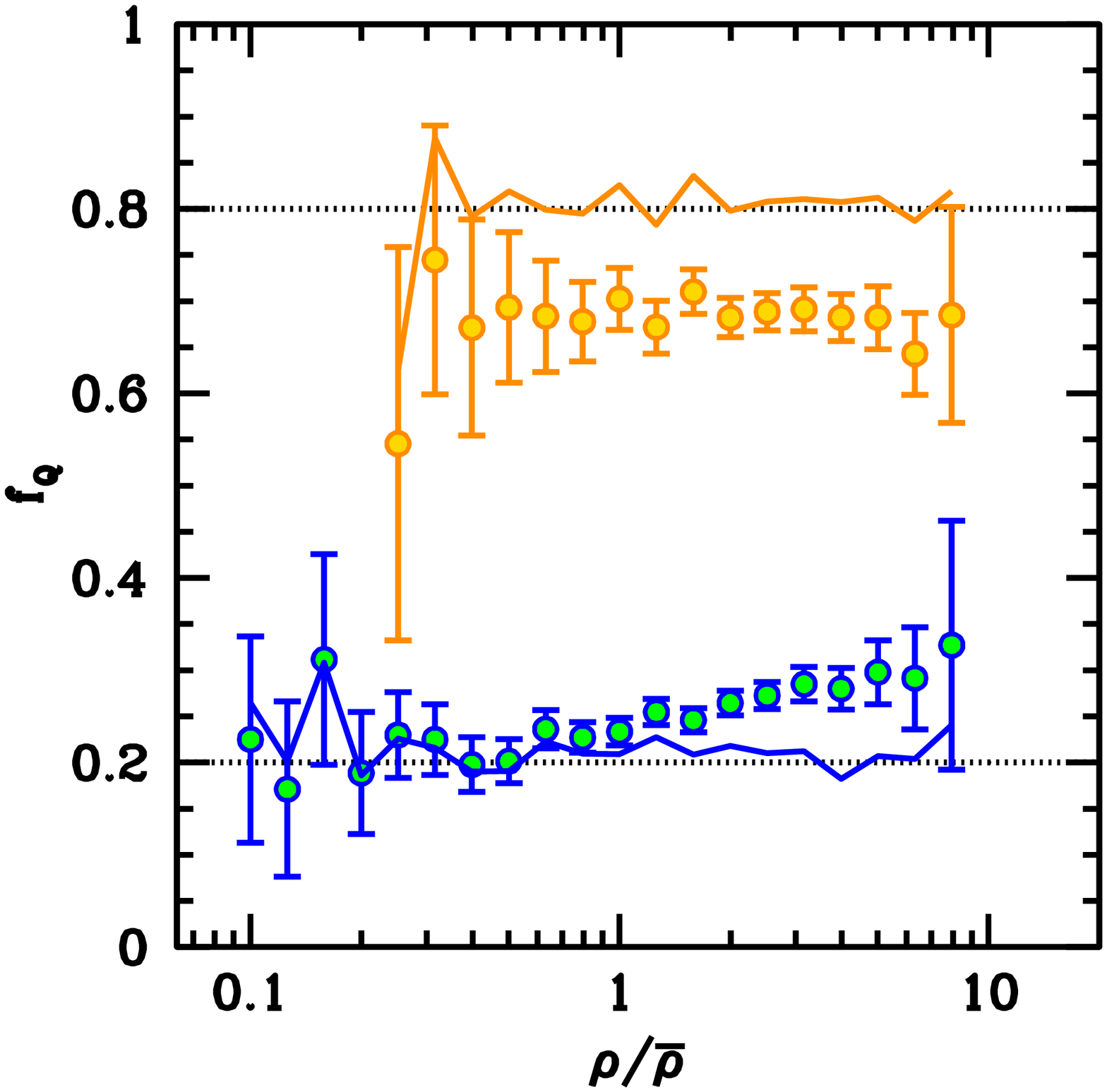,width=0.45\linewidth,clip=} 
\end{tabular}
\caption{ \label{fquench_test} {\it Left panel:} The purity and
  completeness of the group finder vs. halo mass for all galaxies with 
  $M_r < -19$, determined from applying the group finder to a mock galaxy
  distribution from simulation.
  Purity is defined by the fraction of centrals/satellites in the group 
  finder that were of the same type in the input catalog, while 
  completeness is defined by the fraction of centrals/satellites in the input 
  distribution that were assigned to the same type by the group finder. {\it
    Right panel:} Quenched fractions for central and satellite
  galaxies vs. 10 \hmpc\ overdensity.
  Dotted lines show the input values, circles show the measurements from 
  the group finder, and solid curves show the corrected values from equations 
  (\ref{e.fqsat_true}) and (\ref{e.fqsatobs}). }
\end{figure*}

The left panel in Fig.~\ref{fquench_test} shows purity and
completeness of the groups determined via the mock. These statistics
are broken down into central and satellite galaxies, plotted as a
function of (intrinsic) host halo mass.  Across most halo masses, the
purity and completeness of central galaxies is over 90\%. The purity
and completeness of satellite galaxies is lower, $\sim 80\%$. The
average overall fraction of galaxies that are misclassified---labeled 
as a central when they are intrinsically a satellite (or vice versa)---is 
10\%.

To test our recovery of the quenched fraction of galaxies, we randomly
select 20\% of all centrals to be quenched and we randomly select 80\%
of all satellite galaxies to be quenched.  These values were chosen to
make the difference between central and satellite galaxies large in
order to maximize any bias accrued by the group finder. The impurity
and incompleteness of the group finding algorithm yields errors in the
observed quenched fraction of central and satellite galaxies. The
right panel in Fig.~\ref{fquench_test} shows the resulting values of
$\fq$ as a function of density, similar to
Fig.~\ref{fq_den_decomp}. The results of the group finder are shown
with the circles with error bars. The observed $\fqsat$ is reasonably
accurate, but the observed $\fqsat$ is biased low by more than
10\%. At higher densities, the fraction of satellite galaxies
increases, yielding a monotonically increasing $\fqcen$ with
$\rho/\rhobar$.

If the fraction of misclassified galaxies is known, then this bias can
be corrected for statistically. The number of quenched satellite
galaxies is defined as

\begin{equation}
\label{e.fq1}
N_{\rm sat} \fqsatobs =  x\fqsat N_{\rm sat} + (1-y) \fqcen N_{\rm cen},
\end{equation}

\noindent where $x$ is the fraction of satellite galaxies correctly
identified as satellites, and $y$ is the fraction of centrals
correctly identified as centrals. These two quantities are related by

\begin{equation}
\label{e.xy}
(1-x)N_{\rm sat} = (1-y)N_{\rm cen}.
\end{equation}

\noindent Equation (\ref{e.fq1}) reduces to

\begin{equation}
\label{e.fqsatobs}
\fqsatobs = x\fqsat + (1-x) \fqcen,
\end{equation}

\noindent and the observed quenched fraction for centrals can be
similarly written as

\begin{equation}
\label{e.fqcenobs}
\fqcenobs = (1-y)\fqsat + y \fqcen.
\end{equation}

\noindent Solving equation (\ref{e.fqcenobs}) for $\fqcen$ and
substituting into equation (\ref{e.fqsatobs}), the intrinsic $\fqsat$
is given by

\begin{equation}
\label{e.fqsat_true}
\fqsat = \frac{y \fqsatobs + (x-1) \fqcenobs}{xy + (x-1)(1-y)}.
\end{equation}

\noindent Substitution of $\fqsat$ back into
equation(\ref{e.fqcenobs}) yields the equation for $\fqcen$. We find
that $x\approx 0.8$ in our $N$-body simulations (see
Fig.~\ref{fquench_test}, left panel), nearly independent of halo mass
and large-scale density. To determine the overall quenched fraction of
centrals, the misclassified fraction of centrals is $y\approx
0.8N_{\rm sat}/N_{\rm cen}=0.93$. To determine the $\fqcen$ as a
function of density, $y$ is calculated at each value of
$\rho/\rhobar$.  In practice, this procedure overcorrects for $\fqcen$
at the densities where $N_{\rm sat}>N_{\rm cen}$ (roughly
$\rho/\rhobar\gtrsim 6$, or the rightmost 2 data points in
Fig.~\ref{fquench_test}). In this regime, we fix $y=0.85$ and
calculate $x$ through equation (\ref{e.xy}).

The solid curves in right panel of Fig.~\ref{fquench_test} show the
results of applying the above procedure to the measured values of
$\fqcenobs$ and $\fqsatobs$.

\bibliography{../risa}

\label{lastpage}

\end{document}

%% file: newcommands.tex
\def\wp{$w_p(r_p)$}
\def\hmpc{$h^{-1}$Mpc}
\def\hkpc{$h^{-1}$kpc}

\def\msol{M$_\odot$}
\def\hmsol{$h^{-1}$M$_\odot$}

\def\nsat{\langle N_{\rm sat}\rangle_M}
\def\ncen{\langle N_{\rm cen}\rangle_M}

\def\om{\Omega_m}

\def\omb{\Omega_b}
\def\s8{\sigma_8}

\def\lcdm{$\Lambda$CDM}

\def\x2{$\chi^2$}
\def\hmsol{$h^{-1}\,$M$_\odot$}

\def\NNm1{\langle N(N-1) \rangle}

\def\m_star{M_\ast}

\def\rhob{\tilde{\rho}}

\def\lcdm{$\Lambda$CDM}

\def\om{\Omega_m}

\def\omb{\Omega_b}
\def\s8{\sigma_8}

\def\hmpc{$h^{-1}\,$Mpc}
\def\hkpc{$h^{-1}\,$kpc}

\def\x2{$\chi^2$}
\def\hmsol{$h^{-1}\,$M$_\odot$}

\def\wp{$w_p(r_p)$}

\def\nsat{\langle N_{\mbox{\scriptsize sat}}\rangle_M}

\def\ncen{\langle N_{\mbox{\scriptsize cen}}\rangle_M}

\def\NNm1{\langle N(N-1) \rangle}

\def\p0{P_0(r)}

%% file: ms.bbl
\begin{thebibliography}{}

\bibitem[\protect\citeauthoryear{{Abazajian} et~al.,}{{Abazajian}
  et~al.}{2009}]{dr7}
{Abazajian} K.~N.,  et~al., 2009, \apjs, 182, 543

\bibitem[\protect\citeauthoryear{{Abbas} \& {Sheth}}{{Abbas} \&
  {Sheth}}{2006}]{abbas_sheth:06}
{Abbas} U.,  {Sheth} R.~K.,  2006, \mnras, 372, 1749

\bibitem[\protect\citeauthoryear{{Baldry}, {Balogh}, {Bower}, {Glazebrook},
  {Nichol}, {Bamford} \& {Budavari}}{{Baldry} et~al.}{2006}]{baldry_etal:06}
{Baldry} I.~K.,  {Balogh} M.~L.,  {Bower} R.~G.,  {Glazebrook} K.,  {Nichol}
  R.~C.,  {Bamford} S.~P.,    {Budavari} T.,  2006, \mnras, 373, 469

\bibitem[\protect\citeauthoryear{{Balogh} et~al.}{2004}]{balogh_etal:04}
{Balogh} M.,  {Eke} V.,  {Miller} C.,  {Lewis} I.,  {Bower} R.,  {Couch} W.,
  {Nichol} R.,  {Bland-Hawthorn} J.,  {Baldry} I.~K.,  {Baugh} C.,  {Bridges}
  T.,  {Cannon} R.,  {Cole} S.,  {Colless} M.,  {Collins} C.,  {Cross} N.,
  {Dalton} G.,  {de Propris} R.,  {Driver} S.~P.,  {Efstathiou} G.,  {Ellis}
  R.~S.,  {Frenk} C.~S.,  {Glazebrook} K.,  {Gomez} P.,  {Gray} A.,  {Hawkins}
  E.,  {Jackson} C.,  {Lahav} O.,  {Lumsden} S.,  {Maddox} S.,  {Madgwick} D.,
  {Norberg} P.,  {Peacock} J.~A.,  {Percival} W.,  {Peterson} B.~A.,
  {Sutherland} W.,    {Taylor} K.,  2004, \mnras, 348, 1355

\bibitem[\protect\citeauthoryear{{Bamford}, {Nichol}, {Baldry}, {Land},
  {Lintott}, {Schawinski}, {Slosar}, {Szalay}, {Thomas}, {Torki}, {Andreescu},
  {Edmondson}, {Miller}, {Murray}, {Raddick} \& {Vandenberg}}{{Bamford}
  et~al.}{2009}]{bamford_etal:09}
{Bamford} S.~P.,  {Nichol} R.~C.,  {Baldry} I.~K.,  {Land} K.,  {Lintott}
  C.~J.,  {Schawinski} K.,  {Slosar} A.,  {Szalay} A.~S.,  {Thomas} D.,
  {Torki} M.,  {Andreescu} D.,  {Edmondson} E.~M.,  {Miller} C.~J.,  {Murray}
  P.,  {Raddick} M.~J.,    {Vandenberg} J.,  2009, \mnras, 393, 1324

\bibitem[\protect\citeauthoryear{{Behroozi}, {Conroy} \& {Wechsler}}{{Behroozi}
  et~al.}{2010}]{behroozi_etal:10}
{Behroozi} P.~S.,  {Conroy} C.,    {Wechsler} R.~H.,  2010, \apj, 717, 379

\bibitem[\protect\citeauthoryear{{Bell}, {Wolf}, {Meisenheimer}, {Rix},
  {Borch}, {Dye}, {Kleinheinrich}, {Wisotzki} \& {McIntosh}}{{Bell}
  et~al.}{2004}]{bell_etal:04}
{Bell} E.~F.,  {Wolf} C.,  {Meisenheimer} K.,  {Rix} H.-W.,  {Borch} A.,  {Dye}
  S.,  {Kleinheinrich} M.,  {Wisotzki} L.,    {McIntosh} D.~H.,  2004, \apj,
  608, 752

\bibitem[\protect\citeauthoryear{{Berlind}, et~al.}{2006}]{berlind_etal:06_catalog}
{Berlind} A.~A.,  {Frieman} J.,  {Weinberg} D.~H.,  {Blanton} M.~R.,  {Warren}
  M.~S.,  {Abazajian} K.,  {Scranton} R.,  {Hogg} D.~W.,  {Scoccimarro} R.,
  {Bahcall} N.~A.,  {Brinkmann} J.,  {Gott} III J.~R.,  {Kleinman} S.~J.,
  {Krzesinski} J.,  {Lee} B.~C.,  {Miller} C.~J.,  {Nitta} A.,  {Schneider}
  D.~P.,  {Tucker} D.~L.,    {Zehavi} I.,  2006, \apjs, 167, 1

\bibitem[\protect\citeauthoryear{{Berlind} \& {Weinberg}}{{Berlind} \&
  {Weinberg}}{2002}]{berlind_weinberg:02}
{Berlind} A.~A.,  {Weinberg} D.~H.,  2002, \apj, 575, 587

\bibitem[\protect\citeauthoryear{{Blanton} \& {Berlind}}{{Blanton} \&
  {Berlind}}{2007}]{blanton_berlind:07}
{Blanton} M.~R.,  {Berlind} A.~A.,  2007, \apj, 664, 791

\bibitem[\protect\citeauthoryear{{Blanton}, {Eisenstein}, {Hogg}, {Schlegel} \&
  {Brinkmann}}{{Blanton} et~al.}{2005}]{blanton_etal:05a}
{Blanton} M.~R.,  {Eisenstein} D.,  {Hogg} D.~W.,  {Schlegel} D.~J.,
  {Brinkmann} J.,  2005, \apj, 629, 143

\bibitem[\protect\citeauthoryear{{Blanton}, {Eisenstein}, {Hogg} \&
  {Zehavi}}{{Blanton} et~al.}{2006}]{blanton_etal:06a}
{Blanton} M.~R.,  {Eisenstein} D.,  {Hogg} D.~W.,    {Zehavi} I.,  2006, \apj,
  645, 977

\bibitem[\protect\citeauthoryear{{Blanton}, et~al.}{2003}]{blanton_etal:03cmd}
{Blanton} M.~R.,  {Hogg} D.~W.,  {Bahcall} N.~A.,  {Baldry} I.~K.,  {Brinkmann}
  J.,  {Csabai} I.,  {Eisenstein} D.,  {Fukugita} M.,  {Gunn} J.~E.,
  {Ivezi{\'c}} {\v Z}.,  {Lamb} D.~Q.,  {Lupton} R.~H.,  {Loveday} J.,  {Munn}
  J.~A.,  {Nichol} R.~C.,  {Okamura} S.,  {Schlegel} D.~J.,  {Shimasaku} K.,
  {Strauss} M.~A.,  {Vogeley} M.~S.,    {Weinberg} D.~H.,  2003, \apj, 594, 186

\bibitem[\protect\citeauthoryear{{Blanton}, et~al.}{2003}]{blanton_etal:03}
{Blanton} M.~R.,  {Hogg} D.~W.,  {Bahcall} N.~A.,  {Brinkmann} J.,  {Britton}
  M.,  {Connolly} A.~J.,  {Csabai} I.,  {Fukugita} M.,  {Loveday} J.,
  {Meiksin} A.,  {Munn} J.~A.,  {Nichol} R.~C.,  {Okamura} S.,  {Quinn} T.,
  {Schneider} D.~P.,  {Shimasaku} K.,  {Strauss} M.~A.,  {Tegmark} M.,
  {Vogeley} M.~S.,    {Weinberg} D.~H.,  2003, \apj, 592, 819

\bibitem[\protect\citeauthoryear{{Blanton} \& {Roweis}}{{Blanton} \&
  {Roweis}}{2007}]{blanton_roweis:07}
{Blanton} M.~R.,  {Roweis} S.,  2007, \aj, 133, 734

\bibitem[\protect\citeauthoryear{{Blanton}, {Schlegel}, {Strauss}, {Brinkmann},
  {Finkbeiner}, {Fukugita}, {Gunn}, {Hogg}, {Ivezi{\'c}}, {Knapp}, {Lupton},
  {Munn}, {Schneider}, {Tegmark} \& {Zehavi}}{{Blanton}
  et~al.}{2005}]{blanton_etal:05_vagc}
{Blanton} M.~R.,  {Schlegel} D.~J.,  {Strauss} M.~A.,  {Brinkmann} J.,
  {Finkbeiner} D.,  {Fukugita} M.,  {Gunn} J.~E.,  {Hogg} D.~W.,  {Ivezi{\'c}}
  {\v Z}.,  {Knapp} G.~R.,  {Lupton} R.~H.,  {Munn} J.~A.,  {Schneider} D.~P.,
  {Tegmark} M.,    {Zehavi} I.,  2005, \aj, 129, 2562

\bibitem[\protect\citeauthoryear{{Bond}, {Cole}, {Efstathiou} \&
  {Kaiser}}{{Bond} et~al.}{1991}]{bond_etal:91}
{Bond} J.~R.,  {Cole} S.,  {Efstathiou} G.,    {Kaiser} N.,  1991, \apj, 379,
  440

\bibitem[\protect\citeauthoryear{{Bower}, {Benson}, {Malbon}, {Helly}, {Frenk},
  {Baugh}, {Cole} \& {Lacey}}{{Bower} et~al.}{2006}]{bower_etal:06}
{Bower} R.~G.,  {Benson} A.~J.,  {Malbon} R.,  {Helly} J.~C.,  {Frenk} C.~S.,
  {Baugh} C.~M.,  {Cole} S.,    {Lacey} C.~G.,  2006, \mnras, 370, 645

\bibitem[\protect\citeauthoryear{{Brinchmann}, {Charlot}, {White}, {Tremonti},
  {Kauffmann}, {Heckman} \& {Brinkmann}}{{Brinchmann}
  et~al.}{2004}]{brinchmann_etal:04}
{Brinchmann} J.,  {Charlot} S.,  {White} S.~D.~M.,  {Tremonti} C.,  {Kauffmann}
  G.,  {Heckman} T.,    {Brinkmann} J.,  2004, \mnras, 351, 1151

\bibitem[\protect\citeauthoryear{{Brooks}, {Governato}, {Quinn}, {Brook} \&
  {Wadsley}}{{Brooks} et~al.}{2009}]{brooks_etal:09}
{Brooks} A.~M.,  {Governato} F.,  {Quinn} T.,  {Brook} C.~B.,    {Wadsley} J.,
  2009, \apj, 694, 396

\bibitem[\protect\citeauthoryear{{Chabrier}}{{Chabrier}}{2003}]{chabrier:03}
{Chabrier} G.,  2003, \pasp, 115, 763

\bibitem[\protect\citeauthoryear{{Conroy}, {Gunn} \& {White}}{{Conroy}
  et~al.}{2009}]{conroy_etal:09}
{Conroy} C.,  {Gunn} J.~E.,    {White} M.,  2009, \apj, 699, 486

\bibitem[\protect\citeauthoryear{{Conroy} \& {Gunn}}{{Conroy} \&
  {Gunn}}{2010}]{conroy_gunn:10}
{Conroy} C.,  {Gunn} J.~E.,  2010, \apj, 712, 833

\bibitem[\protect\citeauthoryear{{Conroy} \& {Wechsler}}{{Conroy} \&
  {Wechsler}}{2009}]{conroy_wechsler:09}
{Conroy} C.,  {Wechsler} R.~H.,  2009, \apj, 696, 620

\bibitem[\protect\citeauthoryear{{Conroy}, {Wechsler} \& {Kravtsov}}{{Conroy}
  et~al.}{2006}]{conroy_etal:06}
{Conroy} C.,  {Wechsler} R.~H.,    {Kravtsov} A.~V.,  2006, \apj, 647, 201

\bibitem[\protect\citeauthoryear{{Cooper}, {Newman}, {Madgwick}, {Gerke}, {Yan}
  \& {Davis}}{{Cooper} et~al.}{2005}]{cooper_etal:05}
{Cooper} M.~C.,  {Newman} J.~A.,  {Madgwick} D.~S.,  {Gerke} B.~F.,  {Yan} R.,
    {Davis} M.,  2005, \apj, 634, 833

\bibitem[\protect\citeauthoryear{{Cooper}, {Newman}, {Coil}, {Croton}, {Gerke},
  {Yan}, {Davis}, {Faber}, {Guhathakurta}, {Koo}, {Weiner} \&
  {Willmer}}{{Cooper} et~al.}{2007}]{cooper_etal:07}
{Cooper} M.~C.,  {Newman} J.~A.,  {Coil} A.~L.,  {Croton} D.~J.,  {Gerke}
  B.~F.,  {Yan} R.,  {Davis} M.,  {Faber} S.~M.,  {Guhathakurta} P.,  {Koo}
  D.~C.,  {Weiner} B.~J.,    {Willmer} C.~N.~A.,  2007, \mnras, 376, 1445

\bibitem[\protect\citeauthoryear{{Cooper}, {Newman}, {Croton}, {Weiner},
  {Willmer}, {Gerke}, {Madgwick}, {Faber}, {Davis}, {Coil}, {Finkbeiner},
  {Guhathakurta} \& {Koo}}{{Cooper} et~al.}{2006}]{cooper_etal:06}
{Cooper} M.~C.,  {Newman} J.~A.,  {Croton} D.~J.,  {Weiner} B.~J.,  {Willmer}
  C.~N.~A.,  {Gerke} B.~F.,  {Madgwick} D.~S.,  {Faber} S.~M.,  {Davis} M.,
  {Coil} A.~L.,  {Finkbeiner} D.~P.,  {Guhathakurta} P.,    {Koo} D.~C.,  2006,
  \mnras, 370, 198

\bibitem[\protect\citeauthoryear{{Cooray} \& {Sheth}}{{Cooray} \&
  {Sheth}}{2002}]{cooray_sheth:02}
{Cooray} A.,  {Sheth} R.,  2002, \physrep, 372, 1

\bibitem[\protect\citeauthoryear{{Croton} \& {Farrar}}{{Croton} \&
  {Farrar}}{2008}]{croton_farrar:08}
{Croton} D.~J.,  {Farrar} G.~R.,  2008, \mnras, 386, 2285

\bibitem[\protect\citeauthoryear{{Croton}, {Gao} \& {White}}{{Croton}
  et~al.}{2007}]{croton_etal:07}
{Croton} D.~J.,  {Gao} L.,    {White} S.~D.~M.,  2007, \mnras, 374, 1303

\bibitem[\protect\citeauthoryear{{Croton}, {Springel}, {White}, {De Lucia},
  {Frenk}, {Gao}, {Jenkins}, {Kauffmann}, {Navarro} \& {Yoshida}}{{Croton}
  et~al.}{2006}]{croton_etal:06a}
{Croton} D.~J.,  {Springel} V.,  {White} S.~D.~M.,  {De Lucia} G.,  {Frenk}
  C.~S.,  {Gao} L.,  {Jenkins} A.,  {Kauffmann} G.,  {Navarro} J.~F.,
  {Yoshida} N.,  2006, \mnras, 365, 11

\bibitem[\protect\citeauthoryear{{Cucciati} et~al.,}{{Cucciati}
  et~al.}{2006}]{cucciati_etal:06}
{Cucciati} O.,  et~al., 2006, \aap, 458, 39

\bibitem[\protect\citeauthoryear{{Dalal}, {White}, {Bond} \&
  {Shirokov}}{{Dalal} et~al.}{2008}]{dalal_etal:08}
{Dalal} N.,  {White} M.,  {Bond} J.~R.,    {Shirokov} A.,  2008, \apj, 687, 12

\bibitem[\protect\citeauthoryear{{Davis} \& {Geller}}{{Davis} \&
  {Geller}}{1976}]{davis_geller:76}
{Davis} M.,  {Geller} M.~J.,  1976, \apj, 208, 13

\bibitem[\protect\citeauthoryear{{Dekel} \& {Birnboim}}{{Dekel} \&
  {Birnboim}}{2006}]{dekel_birnboim:06}
{Dekel} A.,  {Birnboim} Y.,  2006, \mnras, 368, 2

\bibitem[\protect\citeauthoryear{{Dressler}}{{Dressler}}{1980}]{dressler:80}
{Dressler} A.,  1980, \apj, 236, 351

\bibitem[\protect\citeauthoryear{{Drory}, {Bundy}, {Leauthaud}, {Scoville},
  {Capak}, {Ilbert}, {Kartaltepe}, {Kneib}, {McCracken}, {Salvato}, {Sanders},
  {Thompson} \& {Willott}}{{Drory} et~al.}{2009}]{drory_etal:09}
{Drory} N.,  {Bundy} K.,  {Leauthaud} A.,  {Scoville} N.,  {Capak} P.,
  {Ilbert} O.,  {Kartaltepe} J.~S.,  {Kneib} J.~P.,  {McCracken} H.~J.,
  {Salvato} M.,  {Sanders} D.~B.,  {Thompson} D.,    {Willott} C.~J.,  2009,
  \apj, 707, 1595

\bibitem[\protect\citeauthoryear{{Fakhouri} \& {Ma}}{{Fakhouri} \&
  {Ma}}{2009}]{fakhouri_ma:09}
{Fakhouri} O.,  {Ma} C.-P.,  2009, \mnras, 394, 1825

\bibitem[\protect\citeauthoryear{{Gao}, {Springel} \& {White}}{{Gao}
  et~al.}{2005}]{gao_etal:05}
{Gao} L.,  {Springel} V.,    {White} S.~D.~M.,  2005, MNRAS, 363, L66

\bibitem[\protect\citeauthoryear{{Gao} \& White}{{Gao} \&
  White}{2006}]{gao_white:06}
{Gao} L.,  White S. D.~M., , 2006

\bibitem[\protect\citeauthoryear{{Gill}, {Knebe} \& {Gibson}}{{Gill}
  et~al.}{2005}]{gill_etal:05}
{Gill} S.~P.~D.,  {Knebe} A.,    {Gibson} B.~K.,  2005, \mnras, 356, 1327

\bibitem[\protect\citeauthoryear{{Haas}, {Schaye} \& {Jeeson-Daniel}}{{Haas}
  et~al.}{2011}]{haas_etal:11}
{Haas} M.~R.,  {Schaye} J.,    {Jeeson-Daniel} A.,  2011, \mnras, submitted,
  ArXiv:1103.0547

\bibitem[\protect\citeauthoryear{{Hansen}, {Sheldon}, {Wechsler} \&
  {Koester}}{{Hansen} et~al.}{2009}]{hansen_etal:09}
{Hansen} S.~M.,  {Sheldon} E.~S.,  {Wechsler} R.~H.,    {Koester} B.~P.,  2009,
  \apj, 699, 1333

\bibitem[\protect\citeauthoryear{{Harker}, {Cole}, {Helly}, {Frenk} \&
  {Jenkins}}{{Harker} et~al.}{2006}]{harker_etal:06}
{Harker} G.,  {Cole} S.,  {Helly} J.,  {Frenk} C.,    {Jenkins} A.,  2006,
  \mnras, 367, 1039

\bibitem[\protect\citeauthoryear{{Hogg}, {Blanton}, {Brinchmann}, {Eisenstein},
  {Schlegel}, {Gunn}, {McKay}, {Rix}, {Bahcall}, {Brinkmann} \&
  {Meiksin}}{{Hogg} et~al.}{2004}]{hogg_etal:04}
{Hogg} D.~W.,  {Blanton} M.~R.,  {Brinchmann} J.,  {Eisenstein} D.~J.,
  {Schlegel} D.~J.,  {Gunn} J.~E.,  {McKay} T.~A.,  {Rix} H.,  {Bahcall} N.~A.,
   {Brinkmann} J.,    {Meiksin} A.,  2004, \apjl, 601, L29

\bibitem[\protect\citeauthoryear{{Hopkins}, {Cox}, {Kere{\v s}} \&
  {Hernquist}}{{Hopkins} et~al.}{2008}]{hopkins_etal:08b}
{Hopkins} P.~F.,  {Cox} T.~J.,  {Kere{\v s}} D.,    {Hernquist} L.,  2008,
  \apjs, 175, 390

\bibitem[\protect\citeauthoryear{{Kauffmann}, {Heckman}, {White}, {Charlot},
  {Tremonti}, {Peng}, {Seibert}, {Brinkmann}, {Nichol}, {SubbaRao} \&
  {York}}{{Kauffmann} et~al.}{2003}]{kauffmann_etal:03b}
{Kauffmann} G.,  {Heckman} T.~M.,  {White} S.~D.~M.,  {Charlot} S.,  {Tremonti}
  C.,  {Peng} E.~W.,  {Seibert} M.,  {Brinkmann} J.,  {Nichol} R.~C.,
  {SubbaRao} M.,    {York} D.,  2003, \mnras, 341, 54

\bibitem[\protect\citeauthoryear{{Kauffmann}, {White}, {Heckman}, {M{\'e}nard},
  {Brinchmann}, {Charlot}, {Tremonti} \& {Brinkmann}}{{Kauffmann}
  et~al.}{2004}]{kauffmann_etal:04}
{Kauffmann} G.,  {White} S.~D.~M.,  {Heckman} T.~M.,  {M{\'e}nard} B.,
  {Brinchmann} J.,  {Charlot} S.,  {Tremonti} C.,    {Brinkmann} J.,  2004,
  \mnras, 353, 713

\bibitem[\protect\citeauthoryear{{Kere{\v s}}, {Katz}, {Fardal}, {Dav{\'e}} \&
  {Weinberg}}{{Kere{\v s}} et~al.}{2009}]{keres_etal:09}
{Kere{\v s}} D.,  {Katz} N.,  {Fardal} M.,  {Dav{\'e}} R.,    {Weinberg} D.~H.,
   2009, \mnras, 395, 160

\bibitem[\protect\citeauthoryear{{Kere{\v s}}, {Katz}, {Weinberg} \&
  {Dav{\'e}}}{{Kere{\v s}} et~al.}{2005}]{keres_etal:05}
{Kere{\v s}} D.,  {Katz} N.,  {Weinberg} D.~H.,    {Dav{\'e}} R.,  2005,
  \mnras, 363, 2

\bibitem[\protect\citeauthoryear{{Kimm}, {Somerville}, {Yi}, {van den Bosch},
  {Salim}, {Fontanot}, {Monaco}, {Mo}, {Pasquali}, {Rich} \& {Yang}}{{Kimm}
  et~al.}{2009}]{kimm_etal:09}
{Kimm} T.,  {Somerville} R.~S.,  {Yi} S.~K.,  {van den Bosch} F.~C.,  {Salim}
  S.,  {Fontanot} F.,  {Monaco} P.,  {Mo} H.,  {Pasquali} A.,  {Rich} R.~M.,
  {Yang} X.,  2009, \mnras, 394, 1131

\bibitem[\protect\citeauthoryear{{Kravtsov}, {Berlind}, {Wechsler}, {Klypin},
  {Gottl{\" o}ber}, {Allgood} \& {Primack}}{{Kravtsov}
  et~al.}{2004}]{kravtsov_etal:04}
{Kravtsov} A.~V.,  {Berlind} A.~A.,  {Wechsler} R.~H.,  {Klypin} A.~A.,
  {Gottl{\" o}ber} S.,  {Allgood} B.,    {Primack} J.~R.,  2004, \apj, 609, 35

\bibitem[\protect\citeauthoryear{{Landy} \& {Szalay}}{{Landy} \&
  {Szalay}}{1993}]{landy_szalay:93}
{Landy} S.~D.,  {Szalay} A.~S.,  1993, \apj, 412, 64

\bibitem[\protect\citeauthoryear{{Leauthaud}, {Tinker}, {Behroozi}, {Busha} \&
  {Wechsler}}{{Leauthaud} et~al.}{2011}]{leauthaud_etal:11a}
{Leauthaud} A.,  {Tinker} J.,  {Behroozi} P.~S.,  {Busha} M.~T.,    {Wechsler}
  R.,  2011, ArXiv:1103.2077

\bibitem[\protect\citeauthoryear{{Leauthaud} et~al.}{2011b}]{leauthaud_etal:11b}
{Leauthaud} A.,  {Tinker} J.,  {Bundy} K.,  {Behroozi} P.~S.,  {Massey} R.,
  {Rhodes} J.,  {George} M.~R.,  {Kneib} J.-P.,  {Benson} A.,  {Wechsler}
  R.~H.,  {Busha} M.~T.,  {Capak} P.,  {Cortes} M.,  {Ilbert} O.,  {Koekemoer}
  A.~M.,  {Le Fevre} O.,  {Lilly} S.,  {McCracken} H.~J.,  {Salvato} M.,
  {Schrabback} T.,  {Scoville} N.,  {Smith} T.,    {Taylor} J.~E.,  2011b, ArXiv:1104.0928


\bibitem[\protect\citeauthoryear{{Li}, {Kauffmann}, {Jing}, {White},
  {B{\"o}rner} \& {Cheng}}{{Li} et~al.}{2006}]{li_etal:06}
{Li} C.,  {Kauffmann} G.,  {Jing} Y.~P.,  {White} S.~D.~M.,  {B{\"o}rner} G.,
   {Cheng} F.~Z.,  2006, \mnras, 368, 21

\bibitem[\protect\citeauthoryear{{Li} \& {White}}{{Li} \&
  {White}}{2009}]{li_white:09}
{Li} C.,  {White} S.~D.~M.,  2009, \mnras, 398, 2177

\bibitem[\protect\citeauthoryear{{Li}, {Mo} \& {Gao}}{{Li}
  et~al.}{2008}]{li_etal:08}
{Li} Y.,  {Mo} H.~J.,    {Gao} L.,  2008, \mnras, 389, 1419

\bibitem[\protect\citeauthoryear{{Lintott}, {Schawinski}, {Bamford}, {Slosar},
  {Land}, {Thomas}, {Edmondson}, {Masters}, {Nichol}, {Raddick}, {Szalay},
  {Andreescu}, {Murray} \& {Vandenberg}}{{Lintott}
  et~al.}{2011}]{lintott_etal:11}
{Lintott} C.,  {Schawinski} K.,  {Bamford} S.,  {Slosar} A.,  {Land} K.,
  {Thomas} D.,  {Edmondson} E.,  {Masters} K.,  {Nichol} R.~C.,  {Raddick}
  M.~J.,  {Szalay} A.,  {Andreescu} D.,  {Murray} P.,    {Vandenberg} J.,
  2011, \mnras, 410, 166

\bibitem[\protect\citeauthoryear{{Ludlow}, {Navarro}, {Springel}, {Jenkins},
  {Frenk} \& {Helmi}}{{Ludlow} et~al.}{2009}]{ludlow_etal:09}
{Ludlow} A.~D.,  {Navarro} J.~F.,  {Springel} V.,  {Jenkins} A.,  {Frenk}
  C.~S.,    {Helmi} A.,  2009, \apj, 692, 931

\bibitem[\protect\citeauthoryear{{Macci{\`o}}, {Dutton} \& {van den
  Bosch}}{{Macci{\`o}} et~al.}{2008}]{maccio_etal:08}
{Macci{\`o}} A.~V.,  {Dutton} A.~A.,    {van den Bosch} F.~C.,  2008, \mnras,
  391, 1940

\bibitem[\protect\citeauthoryear{{Madgwick}, {Somerville}, {Lahav} \&
  {Ellis}}{{Madgwick} et~al.}{2003}]{madgwick_etal:03}
{Madgwick} D.~S.,  {Somerville} R.,  {Lahav} O.,    {Ellis} R.,  2003, \mnras,
  343, 871

\bibitem[\protect\citeauthoryear{{Maller}}{{Maller}}{2008}]{maller:08}
{Maller} A.~H.,  2008, in {Funes} J.~G.,  {Corsini} E.~M.,  eds, Astronomical
  Society of the Pacific Conference Series Vol.~396 of Astronomical Society of
  the Pacific Conference Series, {Halo Mergers, Galaxy Mergers, and Why Hubble
  Type Depends on Mass}.
pp 251--+

\bibitem[\protect\citeauthoryear{{Maller}, {Berlind}, {Blanton} \&
  {Hogg}}{{Maller} et~al.}{2009}]{maller_etal:09}
{Maller} A.~H.,  {Berlind} A.~A.,  {Blanton} M.~R.,    {Hogg} D.~W.,  2009,
  \apj, 691, 394

\bibitem[\protect\citeauthoryear{{Mandelbaum}, {Seljak}, {Kauffmann}, {Hirata}
  \& {Brinkmann}}{{Mandelbaum} et~al.}{2006}]{mandelbaum_etal:06_gals}
{Mandelbaum} R.,  {Seljak} U.,  {Kauffmann} G.,  {Hirata} C.~M.,    {Brinkmann}
  J.,  2006, \mnras, 368, 715

\bibitem[\protect\citeauthoryear{{Marchesini}, {van Dokkum}, {Quadri},
  {Rudnick}, {Franx}, {Lira}, {Wuyts}, {Gawiser}, {Christlein} \&
  {Toft}}{{Marchesini} et~al.}{2007}]{marchesini_etal:07}
{Marchesini} D.,  {van Dokkum} P.,  {Quadri} R.,  {Rudnick} G.,  {Franx} M.,
  {Lira} P.,  {Wuyts} S.,  {Gawiser} E.,  {Christlein} D.,    {Toft} S.,  2007,
  \apj, 656, 42

\bibitem[\protect\citeauthoryear{{Masters}, {Mosleh}, {Romer}, {Nichol},
  {Bamford}, {Schawinski}, {Lintott}, {Andreescu}, {Campbell}, {Crowcroft},
  {Doyle}, {Edmondson}, {Murray}, {Raddick}, {Slosar}, {Szalay} \&
  {Vandenberg}}{{Masters} et~al.}{2010}]{masters_etal:10}
{Masters} K.~L.,  {Mosleh} M.,  {Romer} A.~K.,  {Nichol} R.~C.,  {Bamford}
  S.~P.,  {Schawinski} K.,  {Lintott} C.~J.,  {Andreescu} D.,  {Campbell}
  H.~C.,  {Crowcroft} B.,  {Doyle} I.,  {Edmondson} E.~M.,  {Murray} P.,
  {Raddick} M.~J.,  {Slosar} A.,  {Szalay} A.~S.,    {Vandenberg} J.,  2010,
  \mnras, 405, 783

\bibitem[\protect\citeauthoryear{{Masters}, {Nichol}, {Bamford}, {Mosleh},
  {Lintott}, {Andreescu}, {Edmondson}, {Keel}, {Murray}, {Raddick},
  {Schawinski}, {Slosar}, {Szalay}, {Thomas} \& {Vandenberg}}{{Masters}
  et~al.}{2010}]{masters_etal:10_dust}
{Masters} K.~L.,  {Nichol} R.,  {Bamford} S.,  {Mosleh} M.,  {Lintott} C.~J.,
  {Andreescu} D.,  {Edmondson} E.~M.,  {Keel} W.~C.,  {Murray} P.,  {Raddick}
  M.~J.,  {Schawinski} K.,  {Slosar} A.,  {Szalay} A.~S.,  {Thomas} D.,
  {Vandenberg} J.,  2010, \mnras, 404, 792

\bibitem[\protect\citeauthoryear{{More}, {van den Bosch}, {Cacciato}, {Skibba},
  {Mo} \& {Yang}}{{More} et~al.}{2010}]{more_etal:10}
{More} S.,  {van den Bosch} F.~C.,  {Cacciato} M.,  {Skibba} R.,  {Mo} H.~J.,
   {Yang} X.,  2010, \mnras, in press, (arXiv:1003.3203), pp 1464--+

\bibitem[\protect\citeauthoryear{{Moster}, {Somerville}, {Maulbetsch}, {van den
  Bosch}, {Macci{\`o}}, {Naab} \& {Oser}}{{Moster}
  et~al.}{2010}]{moster_etal:09}
{Moster} B.~P.,  {Somerville} R.~S.,  {Maulbetsch} C.,  {van den Bosch} F.~C.,
  {Macci{\`o}} A.~V.,  {Naab} T.,    {Oser} L.,  2010, \apj, 710, 903

\bibitem[\protect\citeauthoryear{{Navarro}, {Frenk} \& {White}}{{Navarro}
  et~al.}{1997a}]{navarro_etal:97}
{Navarro} J.~F.,  {Frenk} C.~S.,    {White} S.~D.~M.,  1997a, ApJ, 490, 493

\bibitem[\protect\citeauthoryear{{Navarro}, {Frenk} \& {White}}{{Navarro}
  et~al.}{1997b}]{nfw:97}
{Navarro} J.~F.,  {Frenk} C.~S.,    {White} S. D.~M.,  1997b, \apj, 490, 493

\bibitem[\protect\citeauthoryear{{Noeske}, {Faber}, {Weiner}, {Koo}, {Primack},
  {Dekel}, {Papovich}, {Conselice}, {Le Floc'h}, {Rieke}, {Coil}, {Lotz},
  {Somerville} \& {Bundy}}{{Noeske} et~al.}{2007}]{noeske_etal:07a}
{Noeske} K.~G.,  {Faber} S.~M.,  {Weiner} B.~J.,  {Koo} D.~C.,  {Primack}
  J.~R.,  {Dekel} A.,  {Papovich} C.,  {Conselice} C.~J.,  {Le Floc'h} E.,
  {Rieke} G.~H.,  {Coil} A.~L.,  {Lotz} J.~M.,  {Somerville} R.~S.,    {Bundy}
  K.,  2007, \apjl, 660, L47

\bibitem[\protect\citeauthoryear{{Norberg}, et~al.}{2002}]{norberg_etal:02}
{Norberg} P.,  {Baugh} C.~M.,  {Hawkins} E.,  {Maddox} S.,  {Madgwick} D.,
  {Lahav} O.,  {Cole} S.,  {Frenk} C.~S.,  {Baldry} I.,  {Bland-Hawthorn} J.,
  {Bridges} T.,  {Cannon} R.,  {Colless} M.,  {Collins} C.,  {Couch} W.,
  {Dalton} G.,  {De Propris} R.,  {Driver} S.~P.,  {Efstathiou} G.,  {Ellis}
  R.~S.,  {Glazebrook} K.,  {Jackson} C.,  {Lewis} I.,  {Lumsden} S.,
  {Peacock} J.~A.,  {Peterson} B.~A.,  {Sutherland} W.,    {Taylor} K.,  2002,
  \mnras, 332, 827

\bibitem[\protect\citeauthoryear{{Norberg}, et~al.}{2001}]{norberg_etal:01}
{Norberg} P.,  {Baugh} C.~M.,  {Hawkins} E.,  {Maddox} S.,  {Peacock} J.~A.,
  {Cole} S.,  {Frenk} C.~S.,  {Bland-Hawthorn} J.,  {Bridges} T.,  {Cannon} R.,
   {Colless} M.,  {Collins} C.,  {Couch} W.,  {Dalton} G.,  {De Propris} R.,
  {Driver} S.~P.,  {Efstathiou} G.,  {Ellis} R.~S.,  {Glazebrook} K.,
  {Jackson} C.,  {Lahav} O.,  {Lewis} I.,  {Lumsden} S.,  {Madgwick} D.,
  {Peterson} B.~A.,  {Sutherland} W.,    {Taylor} K.,  2001, \mnras, 328, 64

\bibitem[\protect\citeauthoryear{{Oemler} Jr.}{{Oemler}}{1974}]{oemler:74}
{Oemler} Jr. A.,  1974, \apj, 194, 1

\bibitem[\protect\citeauthoryear{{Park}, {Choi}, {Vogeley}, {Gott} \&
  {Blanton}}{{Park} et~al.}{2007}]{park_etal:07}
{Park} C.,  {Choi} Y.-Y.,  {Vogeley} M.~S.,  {Gott} J.~R.~I.,    {Blanton}
  M.~R.,  2007, \apj, 658, 898

\bibitem[\protect\citeauthoryear{{Peacock} \& {Smith}}{{Peacock} \&
  {Smith}}{2000}]{peacock_smith:00}
{Peacock} J.~A.,  {Smith} R.~E.,  2000, \mnras, 318, 1144


\bibitem[\protect\citeauthoryear{{Peng}, {Lilly}, {Renzini} \&
  {Carollo}}{{Peng} et~al.}{2011}]{peng_etal:11}
{Peng} Y.,  {Lilly} S.~J.,  {Renzini} A.,    {Carollo} M.,  2011, ArXiv:1106.2546

\bibitem[\protect\citeauthoryear{{Scoccimarro}, {Sheth}, {Hui} \&
  {Jain}}{{Scoccimarro} et~al.}{2001}]{roman_etal:01}
{Scoccimarro} R.,  {Sheth} R.~K.,  {Hui} L.,    {Jain} B.,  2001, \apj, 546, 20

\bibitem[\protect\citeauthoryear{{Seljak}}{{Seljak}}{2000}]{seljak:00}
{Seljak} U.,  2000, \mnras, 318, 203

\bibitem[\protect\citeauthoryear{{Sheth} \& {Tormen}}{{Sheth} \&
  {Tormen}}{2002}]{sheth_tormen:02}
{Sheth} R.~K.,  {Tormen} G.,  2002, \mnras, 329, 61

\bibitem[\protect\citeauthoryear{{Skibba}, {Sheth}, {Connolly} \&
  {Scranton}}{{Skibba} et~al.}{2006}]{skibba_etal:06}
{Skibba} R.,  {Sheth} R.~K.,  {Connolly} A.~J.,    {Scranton} R.,  2006,
  \mnras, 369, 68

\bibitem[\protect\citeauthoryear{{Skibba}, {van den Bosch}, {Yang}, {More},
  {Mo} \& {Fontanot}}{{Skibba} et~al.}{2011}]{skibba_etal:11}
{Skibba} R.~A.,  {van den Bosch} F.~C.,  {Yang} X.,  {More} S.,  {Mo} H.,
  {Fontanot} F.,  2011, \mnras, 410, 417

\bibitem[\protect\citeauthoryear{{Somerville}, {Hopkins}, {Cox}, {Robertson} \&
  {Hernquist}}{{Somerville} et~al.}{2008}]{somerville_etal:08}
{Somerville} R.~S.,  {Hopkins} P.~F.,  {Cox} T.~J.,  {Robertson} B.~E.,
  {Hernquist} L.,  2008, \mnras, pp 1241--+

\bibitem[\protect\citeauthoryear{{Strateva}, et~al.}{2001}]{strateva_etal:01}
{Strateva} I.,  {Ivezi{\'c}} {\v Z}.,  {Knapp} G.~R.,  {Narayanan} V.~K.,
  {Strauss} M.~A.,  {Gunn} J.~E.,  {Lupton} R.~H.,  {Schlegel} D.,  {Bahcall}
  N.~A.,  {Brinkmann} J.,  {Brunner} R.~J.,  {Budav{\'a}ri} T.,  {Csabai} I.,
  {Castander} F.~J.,  {Doi} M.,  {Fukugita} M.,  {Gy{\H o}ry} Z.,  {Hamabe} M.,
   {Hennessy} G.,  {Ichikawa} T.,  {Kunszt} P.~Z.,  {Lamb} D.~Q.,  {McKay}
  T.~A.,  {Okamura} S.,  {Racusin} J.,  {Sekiguchi} M.,  {Schneider} D.~P.,
  {Shimasaku} K.,    {York} D.,  2001, \aj, 122, 1861

\bibitem[\protect\citeauthoryear{{Swanson}, {Tegmark}, {Blanton} \&
  {Zehavi}}{{Swanson} et~al.}{2008}]{swanson_etal:08_bias}
{Swanson} M.~E.~C.,  {Tegmark} M.,  {Blanton} M.,    {Zehavi} I.,  2008,
  \mnras, 385, 1635

\bibitem[\protect\citeauthoryear{{Swanson}, {Tegmark}, {Hamilton} \&
  {Hill}}{{Swanson} et~al.}{2008}]{swanson_etal:08}
{Swanson} M.~E.~C.,  {Tegmark} M.,  {Hamilton} A.~J.~S.,    {Hill} J.~C.,
  2008, \mnras, 387, 1391

\bibitem[\protect\citeauthoryear{{Tinker}, {Kravtsov}, {Klypin}, {Abazajian},
  {Warren}, {Yepes}, {Gottl{\"o}ber} \& {Holz}}{{Tinker}
  et~al.}{2008}]{tinker_etal:08_mf}
{Tinker} J.,  {Kravtsov} A.~V.,  {Klypin} A.,  {Abazajian} K.,  {Warren} M.,
  {Yepes} G.,  {Gottl{\"o}ber} S.,    {Holz} D.~E.,  2008, \apj, 688, 709

\bibitem[\protect\citeauthoryear{{Tinker}, {Conroy}, {Norberg}, {Patiri},
  {Weinberg} \& {Warren}}{{Tinker} et~al.}{2008}]{tinker_etal:08_voids}
{Tinker} J.~L.,  {Conroy} C.,  {Norberg} P.,  {Patiri} S.~G.,  {Weinberg}
  D.~H.,    {Warren} M.~S.,  2008, \apj, 686, 53

\bibitem[\protect\citeauthoryear{{Tinker}, {Weinberg}, {Zheng} \&
  {Zehavi}}{{Tinker} et~al.}{2005}]{tinker_etal:05}
{Tinker} J.~L.,  {Weinberg} D.~H.,  {Zheng} Z.,    {Zehavi} I.,  2005, \apj,
  631, 41

\bibitem[\protect\citeauthoryear{{Tinker} \& {Wetzel}}{{Tinker} \&
  {Wetzel}}{2010}]{tinker_wetzel:10}
{Tinker} J.~L.,  {Wetzel} A.~R.,  2010, \apj, 719, 88

\bibitem[\protect\citeauthoryear{{Vale} \& {Ostriker}}{{Vale} \&
  {Ostriker}}{2006}]{vale_ostriker:06}
{Vale} A.,  {Ostriker} J.~P.,  2006, \mnras, 371, 1173

\bibitem[\protect\citeauthoryear{{van den Bosch}, {Aquino}, {Yang}, {Mo},
  {Pasquali}, {McIntosh}, {Weinmann} \& {Kang}}{{van den Bosch}
  et~al.}{2008}]{vdb_etal:08}
{van den Bosch} F.~C.,  {Aquino} D.,  {Yang} X.,  {Mo} H.~J.,  {Pasquali} A.,
  {McIntosh} D.~H.,  {Weinmann} S.~M.,    {Kang} X.,  2008, \mnras, 387, 79

\bibitem[\protect\citeauthoryear{{van den Bosch}, {Pasquali}, {Yang}, {Mo},
  {Weinmann}, {McIntosh} \& {Aquino}}{{van den Bosch}
  et~al.}{2008}]{vdb_etal:08b}
{van den Bosch} F.~C.,  {Pasquali} A.,  {Yang} X.,  {Mo} H.~J.,  {Weinmann} S.,
   {McIntosh} D.~H.,    {Aquino} D.,  2008, \mnras, submitted, ArXiv:0805.0002

\bibitem[\protect\citeauthoryear{{von der Linden}, {Wild}, {Kauffmann}, {White}
  \& {Weinmann}}{{von der Linden} et~al.}{2010}]{vonderlinden_etal:10}
{von der Linden} A.,  {Wild} V.,  {Kauffmann} G.,  {White} S.~D.~M.,
  {Weinmann} S.,  2010, \mnras, 404, 1231

\bibitem[\protect\citeauthoryear{{Wang}, {Li}, {Kauffmann} \& {De
  Lucia}}{{Wang} et~al.}{2006}]{wang_etal:06}
{Wang} L.,  {Li} C.,  {Kauffmann} G.,    {De Lucia} G.,  2006, \mnras, 371, 537

\bibitem[\protect\citeauthoryear{{Wang}, {Yang}, {Mo}, {van den Bosch},
  {Weinmann} \& {Chu}}{{Wang} et~al.}{2008}]{wang_etal:08_assembly_bias}
{Wang} Y.,  {Yang} X.,  {Mo} H.~J.,  {van den Bosch} F.~C.,  {Weinmann} S.~M.,
    {Chu} Y.,  2008, \apj, 687, 919

\bibitem[\protect\citeauthoryear{{Wang}, {Mo} \& {Jing}}{{Wang}
  et~al.}{2009}]{wang_etal:09}
{Wang} H.,  {Mo} H.~J.,    {Jing} Y.~P.,  2009, \mnras, 396, 2249

\bibitem[\protect\citeauthoryear{{Wang}, {Yang}, {Mo}, {van den Bosch}, {Katz},
  {Pasquali}, {McIntosh} \& {Weinmann}}{{Wang}
  et~al.}{2009}]{wang_etal:09_dwarfs}
{Wang} Y.,  {Yang} X.,  {Mo} H.~J.,  {van den Bosch} F.~C.,  {Katz} N.,
  {Pasquali} A.,  {McIntosh} D.~H.,    {Weinmann} S.~M.,  2009, \apj, 697, 247

\bibitem[\protect\citeauthoryear{{Wechsler}, {Zentner}, {Bullock}, {Kravtsov}
  \& {Allgood}}{{Wechsler} et~al.}{2006}]{wechsler_etal:06}
{Wechsler} R.~H.,  {Zentner} A.~R.,  {Bullock} J.~S.,  {Kravtsov} A.~V.,
  {Allgood} B.,  2006, \apj, 652, 71

\bibitem[\protect\citeauthoryear{{Weinmann}, {van den Bosch}, {Yang} \&
  {Mo}}{{Weinmann} et~al.}{2006}]{weinmann_etal:06a}
{Weinmann} S.~M.,  {van den Bosch} F.~C.,  {Yang} X.,    {Mo} H.~J.,  2006,
  \mnras, 366, 2

\bibitem[\protect\citeauthoryear{{Weinmann}, {van den Bosch}, {Yang}, {Mo},
  {Croton} \& {Moore}}{{Weinmann} et~al.}{2006}]{weinmann_etal:06b}
{Weinmann} S.~M.,  {van den Bosch} F.~C.,  {Yang} X.,  {Mo} H.~J.,  {Croton}
  D.~J.,    {Moore} B.,  2006, \mnras, 372, 1161

\bibitem[\protect\citeauthoryear{{Weinmann}, {Kauffmann}, {von der Linden} \&
  {De Lucia}}{{Weinmann} et~al.}{2010}]{weinmann_etal:10}
{Weinmann} S.~M.,  {Kauffmann} G.,  {von der Linden} A.,    {De Lucia} G.,
  2010, \mnras, 406, 2249

\bibitem[\protect\citeauthoryear{{Wetzel}, {Cohn}, {White}, {Holz} \&
  {Warren}}{{Wetzel} et~al.}{2007}]{wetzel_etal:07}
{Wetzel} A.~R.,  {Cohn} J.~D.,  {White} M.,  {Holz} D.~E.,    {Warren} M.~S.,
  2007, \apj, 656, 139

\bibitem[\protect\citeauthoryear{{Wetzel} \& {White}}{{Wetzel} \&
  {White}}{2010}]{wetzel_white:10}
{Wetzel} A.~R.,  {White} M.,  2010, \mnras, 403, 1072

\bibitem[\protect\citeauthoryear{{White}, {Cohn} \& {Smit}}{{White}
  et~al.}{2010}]{white_etal:10}
{White} M.,  {Cohn} J.~D.,    {Smit} R.,  2010, \mnras, 408, 1818

\bibitem[\protect\citeauthoryear{{Williams}, {Quadri}, {Franx}, {van Dokkum} \&
  {Labb{\'e}}}{{Williams} et~al.}{2009}]{williams_etal:09}
{Williams} R.~J.,  {Quadri} R.~F.,  {Franx} M.,  {van Dokkum} P.,
  {Labb{\'e}} I.,  2009, \apj, 691, 1879

\bibitem[\protect\citeauthoryear{{Willmer}, et~al.}{2006}]{willmer_etal:06}
{Willmer} C.~N.~A.,  {Faber} S.~M.,  {Koo} D.~C.,  {Weiner} B.~J.,  {Newman}
  J.~A.,  {Coil} A.~L.,  {Connolly} A.~J.,  {Conroy} C.,  {Cooper} M.~C.,
  {Davis} M.,  {Finkbeiner} D.~P.,  {Gerke} B.~F.,  {Guhathakurta} P.,
  {Harker} J.,  {Kaiser} N.,  {Kassin} S.,  {Konidaris} N.~P.,  {Lin} L.,
  {Luppino} G.,  {Madgwick} D.~S.,  {Noeske} K.~G.,  {Phillips} A.~C.,    {Yan}
  R.,  2006, \apj, 647, 853

\bibitem[\protect\citeauthoryear{{Wilman}, {Zibetti} \&
  {Budav{\'a}ri}}{{Wilman} et~al.}{2010}]{wilman_etal:10}
{Wilman} D.~J.,  {Zibetti} S.,    {Budav{\'a}ri} T.,  2010, \mnras, 406, 1701

\bibitem[\protect\citeauthoryear{{Yang}, {Mo} \& {van den Bosch}}{{Yang}
  et~al.}{2006}]{yang_etal:06}
{Yang} X.,  {Mo} H.~J.,    {van den Bosch} F.~C.,  2006, \apjl, 638, L55

\bibitem[\protect\citeauthoryear{{Yang}, {Mo} \& {van den Bosch}}{{Yang}
  et~al.}{2008}]{yang_etal:08}
{Yang} X.,  {Mo} H.~J.,    {van den Bosch} F.~C.,  2008, \apj, 676, 248

\bibitem[\protect\citeauthoryear{{Yang}, {Mo} \& {van den Bosch}}{{Yang}
  et~al.}{2009}]{yang_etal:09_groups3}
{Yang} X.,  {Mo} H.~J.,    {van den Bosch} F.~C.,  2009, \apj, 695, 900

\bibitem[\protect\citeauthoryear{{Yang}, {Mo}, {van den Bosch} \&
  {Jing}}{{Yang} et~al.}{2005}]{yang_etal:05}
{Yang} X.,  {Mo} H.~J.,  {van den Bosch} F.~C.,    {Jing} Y.~P.,  2005, \mnras,
  356, 1293

\bibitem[\protect\citeauthoryear{{Yang}, {Mo}, {van den Bosch}, {Pasquali},
  {Li} \& {Barden}}{{Yang} et~al.}{2007}]{yang_etal:07_catalog}
{Yang} X.,  {Mo} H.~J.,  {van den Bosch} F.~C.,  {Pasquali} A.,  {Li} C.,
  {Barden} M.,  2007, \apj, 671, 153

\bibitem[\protect\citeauthoryear{{York} et~al.,}{{York}
  et~al.}{2000}]{york_etal:00}
{York} D.~G.,  et~al., 2000, \aj, 120, 1579

\bibitem[\protect\citeauthoryear{{Zehavi}, et~al.,}{{Zehavi} et~al.}{2002}]{zehavi_etal:02}
{Zehavi} I.,  {Blanton} M.~R.,  {Frieman} J.~A.,  {Weinberg} D.~H.,  {Mo}
  H.~J.,  {Strauss} M.~A.,    et~al., 2002, \apj, 571, 172

\bibitem[\protect\citeauthoryear{{Zehavi}, {Zheng}, {Weinberg}, {Blanton},
  {Bahcall}, {Berlind}, {Brinkmann}, {Frieman}, {Gunn}, {Lupton}, {Nichol},
  {Percival}, {Schneider}, {Skibba}, {Strauss}, {Tegmark} \& {York}}{{Zehavi}
  et~al.}{2010}]{zehavi_etal:10}
{Zehavi} I.,  {Zheng} Z.,  {Weinberg} D.~H.,  {Blanton} M.~R.,  {Bahcall}
  N.~A.,  {Berlind} A.~A.,  {Brinkmann} J.,  {Frieman} J.~A.,  {Gunn} J.~E.,
  {Lupton} R.~H.,  {Nichol} R.~C.,  {Percival} W.~J.,  {Schneider} D.~P.,
  {Skibba} R.~A.,  {Strauss} M.~A.,  {Tegmark} M.,    {York} D.~G.,  2010,
  \apj, submitted (arXiv:1005.2413)

\bibitem[\protect\citeauthoryear{{Zehavi}, et~al.}{2005}]{zehavi_etal:05}
{Zehavi} I.,  {Zheng} Z.,  {Weinberg} D.~H.,  {Frieman} J.~A.,  {Berlind}
  A.~A.,  {Blanton} M.~R.,  {Scoccimarro} R.,  {Sheth} R.~K.,  {Strauss} M.~A.,
   {Kayo} I.,  {Suto} Y.,  {Fukugita} M.,  {Nakamura} O.,  {Bahcall} N.~A.,
  {Brinkmann} J.,  {Gunn} J.~E.,  {Hennessy} G.~S.,  {Ivezi{\'c}} {\v Z}.,
  {Knapp} G.~R.,  {Loveday} J.,  {Meiksin} A.,  {Schlegel} D.~J.,  {Schneider}
  D.~P.,  {Szapudi} I.,  {Tegmark} M.,  {Vogeley} M.~S.,    {York} D.~G.,
  2005, \apj, 630, 1

\bibitem[\protect\citeauthoryear{{Zheng}, {Berlind}, {Weinberg}, {Benson},
  {Baugh}, {Cole}, {Dav{\'e}}, {Frenk}, {Katz} \& {Lacey}}{{Zheng}
  et~al.}{2005}]{zheng_etal:05}
{Zheng} Z.,  {Berlind} A.~A.,  {Weinberg} D.~H.,  {Benson} A.~J.,  {Baugh}
  C.~M.,  {Cole} S.,  {Dav{\'e}} R.,  {Frenk} C.~S.,  {Katz} N.,    {Lacey}
  C.~G.,  2005, \apj, 633, 791

\bibitem[\protect\citeauthoryear{{Zhu}, {Zheng}, {Lin}, {Jing}, {Kang} \&
  {Gao}}{{Zhu} et~al.}{2006}]{zhu_etal:06}
{Zhu} G.,  {Zheng} Z.,  {Lin} W.~P.,  {Jing} Y.~P.,  {Kang} X.,    {Gao} L.,
  2006, \apjl, 639, L5

\end{thebibliography}
